\documentclass{aa}
\usepackage{graphics}

\begin{document}

\thesaurus{06
          (11.01.2;
           11.01.2;
           11.09.1;
           11.19.1;
           13.25.2)}

\title{The Warm Absorber constrained by the coronal lines in Seyfert 1 galaxies}
\author{D. Porquet\inst{1}, A.-M. Dumont\inst{1}, S. Collin\inst{1}, and M. Mouchet\inst{1,2}}

\offprints{D.Porquet}
\mail{Delphine.Porquet@obspm.fr}

\institute{DAEC, Observatoire de Paris, Section Meudon, F-92195 Meudon Cedex, France 
\and Universit{\'e} Denis Diderot, F-75251 Paris Cedex 05, France}
\date{Received ; accepted }

\titlerunning{Warm Absorber constrained by the coronal lines in Seyfert 1s}
\authorrunning{Porquet et al.}
\maketitle

\begin{abstract}
We present results of the photoionization code IRIS, which calculates the spectrum emitted by the Warm Absorber (WA) in Seyfert 1 galaxies for a large grid of parameters (density, column density, ionization parameter...). We show that in Seyfert 1s, coronal lines ([\ion{Fe}{x}], [\ion{Fe}{xi}], [\ion{Fe}{xiv}]...), whose emission region shares common characteristics with the WA, could be formed in the WA.
Unlike the absorption edges, such as those of \ion{O}{vii} and \ion{O}{viii} observed in soft X-rays which are produced by the WA, these lines strongly constrain the physical parameters of the WA, especially the hydrogen density. Indeed, in order to avoid producing coronal line equivalent widths larger than observed, a high density (n$_{H} \geq$ 10$^{10}$ cm$^{-3}$) is required for the WA in most cases. This result is obtained for the mean observed Seyfert 1 features, as well as for the case study of \object{MCG-6-30-15}. It implies that the distance of the WA from the incident radiation source is of the order of that of the Broad Line Region (BLR).\\ 
\end{abstract}

\section{Introduction} \label{sec:Introduction}

\indent The Warm Absorber (WA) is an optically thin ionized medium, first proposed by Halpern (\cite{Halpern}) in order to explain the shape of the X-ray spectrum of the \object{QSO MR2251-178} observed with the Einstein Observatory. 
The main signatures of this medium are the two high-ionization oxygen absorption edges, \ion{O}{vii} and \ion{O}{viii} at 0.74 keV and 0.87 keV respectively, seen in about fifty percent of Seyfert 1 galaxies (Nandra $\&$ Pounds \cite{Nandra}; Reynolds 1997, hereafter referred as \cite{Reynolds}).\\
\indent Mihara et al. (\cite{Mihara}) found with ASCA observations of \object{NGC 4051} that the absorption edges of \ion{O}{vii} and \ion{O}{viii} may be blueshifted by about 3$\%$. This could be due to an outflow velocity of $\sim$ 10\,000 km.s$^{-1}$.\\
\indent According to Netzer (\cite{Netzer93}), an emission line spectrum from the WA should also be observed. Indeed, an \ion{O}{vii} line (0.57 keV) was detected in \object{NGC 3783} (George et al. \cite{George}). Other Seyferts may also show oxygen emission lines: \object{MCG-6-30-15} (\ion{O}{vii}-\ion{O}{viii}: Otani et al. 1996 but the authors mentioned that those features could have an instrumental origin), and \object{1E 1615+061} (\ion{O}{vii}-\ion{O}{viii}: Piro et al. \cite{Piro}). The WA may also contribute to the emission of the UV lines (Shields et al. \cite{Shields}, Netzer \cite{Netzer96}), like \ion{Ne}{viii} 774\AA$~$ (Hamann et al. {\cite{Hamann}) and \ion{O}{vi} 1034\AA$~$ which are also produced in highly ionized regions.\\ 
\indent The WA is generally thought to be a photoionized medium which lies on the line of sight of the ionizing X-ray source. But the possibility of collisional ionization is not ruled out, and it is therefore also investigated in this article.\\ 

\indent In Seyfert 1 spectra, coronal lines are also observed. They are fine structure transitions in the ground level of highly ionized ions which have threshold energies above 100 eV.
According to Penston et al. (\cite{Penston}), [\ion{Fe}{x}] 6375\AA$~$ is found in half Seyfert objects (with no preference between type-1 and type-2 Seyferts). From their Table 4, [\ion{Fe}{xi}] 7892\AA$~$ is detected in 6 of their 19 Seyfert 1s. In the sample used by Erkens et al. (1997) (hereafter referred as \cite{Erkens}), the mean widths of the forbidden high ionization lines (FHILs) are intermediate between those of the Broad Line Region (BLR) and the Narrow Line Region (NLR), but in some cases the wings of the FHILs could be comparable to or broader than the BLR profiles. The FHIL region seems therefore to be located near the BLR, or between the BLR and the NLR. According to \cite{Erkens}, it is located outside the BLR if the Unified Scheme is correct, since broad FHILs are also found in Seyfert 2 spectra.
\cite{Erkens} confirmed that FHILs like [\ion{Fe}{x}] 6375\AA, [\ion{Fe}{xi}] 7892\AA$~$ and [\ion{Fe}{xiv}] 5303\AA$~$ are in average broader and more blueshifted than lower ionization lines like [\ion{Ne}{v}] 3426\AA$~$ and [\ion{Fe}{vii}] 6087\AA. The line widths and blueshifts are correlated with the ionization potential; In other words, the more ionized species have the larger outflowing velocities.\\

\indent Therefore, the WA and the high-ionization coronal line region seem to have common characteristics: a high ionization state, a location between the NLR and the BLR, and an outflowing gas.\\
This leads us to discuss the possibility that the coronal lines and the WA features could be produced in the same medium. We shall study the constraints that the coronal line intensities impose on the WA. Preliminary results have been already presented by Porquet $\&$ Dumont (\cite{Porquet}).\\

\indent In Sect.~\ref{sec:Observation}, we report the observational data relative to the coronal lines, to the optical depths of \ion{O}{vii} and \ion{O}{viii}, and to the UV resonance emission lines. In Sect.~\ref{sec:Calculation}, we discuss previous models and describe our computations. The results of a pure photoionized model and of an hybrid model (photoionized medium out of thermal equilibrium) for two shapes of incident continua are given in Sect.~\ref{sec:results} and compared to the mean observed Seyfert 1 features. The particular case of \object{MCG-6-30-15} is treated in Sect.~\ref{sec:mcg6-30-15}. In Sect.~\ref{sec:Conclusion}, we discuss some implications of the results.\\
\indent Throughout this article, we assume H$_{o}$=50 km\,s$^{-1}$\,
Mpc$^{-1}$ and q$_{o}$=0.

\section{Observational data} \label{sec:Observation}

For Figures \ref{f1} to \ref{f5}, when they exist, error bars are reported.

\subsection{Optical depths of Oxygen} \label{sec:tau}

\begin{figure}
\resizebox{7.75cm}{!}{\includegraphics{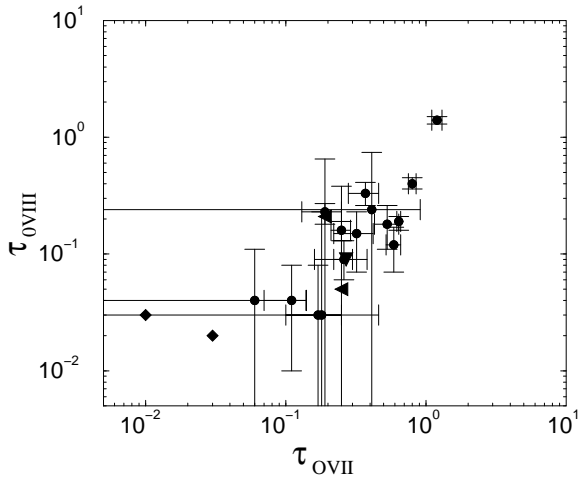}} 
\caption{$\tau_{\ion{O}{viii}}$ versus $\tau_{\ion{O}{vii}}$ for 20 Seyfert 1s taken from Reynolds (\cite{Reynolds97}). {\it Filled circle}: real value, {\it triangle left}: upper limit for $\tau_{\ion{O}{vii}}$, {\it triangle down}: upper limit for $\tau_{\ion{O}{viii}}$ and {\it diamond}: upper limit for both values.}  
\label{f1}
\end{figure}

\begin{figure}
\resizebox{7.75cm}{!}{\includegraphics{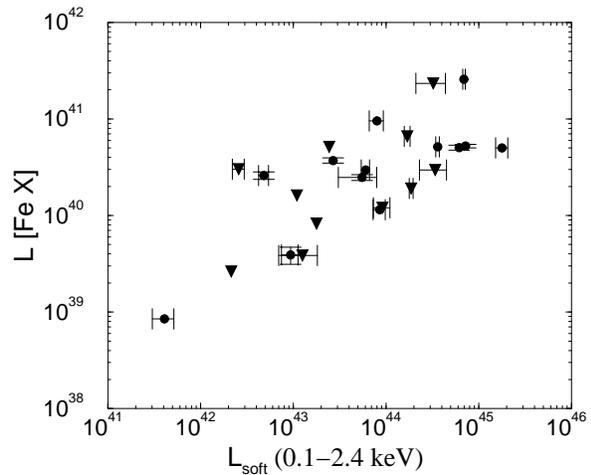}}
\caption{Observed dereddened luminosity of [\ion{Fe}{x}] (erg.s$^{-1}$) versus the soft X-ray luminosity (erg.s$^{-1}$). {\it Filled triangle down}: upper limit for [\ion{Fe}{x}] and {\it filled circle}: real value. [\ion{Fe}{x}] data are from Reynolds et al. (\cite{Reynolds97}) (\object{MCG-6-30-15}), Penston et al. (\cite{Penston}) (\object{NGC 7469}, \object{NGC 4051}, \object{NGC 5548}, \object{NGC 3516}, \object{Mrk 335}, \object{Mrk 509}, \object{IC 4329A}, \object{Fairall 9}, \object{Mrk 79}, \object{ESO G141-55}, \object{NGC 6814}, \object{IZw1}, \object{Mrk 618}, \object{Mrk 926}), \cite{Erkens} (\object{Mrk 9}, \object{Mrk 704}, \object{Mrk 1239}, \object{Akn 120}, \object{Akn 564}), Morris $\&$ Ward (\cite{Morris}) (\object{NGC 4593}, \object{Mrk 1347}, \object{Fairall 51}, \object{UGC 10683B}, but line has not been corrected for blending with [\ion{O}{i}] 6364\AA). 
The soft X-ray luminosities are from Rush et al. (\cite{Rush})  (\object{MCG-6-30-15}, \object{NGC 7469}, \object{NGC 4051}, \object{NGC 5548}, \object{NGC 3516}, \object{Mrk 335}, \object{Mrk 509}, \object{NGC 4593}, \object{IC 4329A}, \object{Mrk 704}, \object{Mrk 79}, \object{Mrk 1239}, \object{ESO G141-55}, \object{IZw1}, \object{Mrk 618}), Schartel et al. (\cite{Schartel}) (\object{Fairall 9}, \object{Mrk 926}), Boller et al. (\cite{Boller}) (\object{Mrk 9}, \object{Akn 120}, \object{Akn 564}, \object{NGC 6814}, \object{Mrk 1347}, \object{Fairall 51}, \object{UGC 10683B}).}
\label{f2}
\end{figure}

\indent All the optical depth values at the \ion{O}{vii} edge ($\tau_{\ion{O}{vii}}$) and \ion{O}{viii} edge ($\tau_{\ion{O}{viii}}$) for Seyfert 1s are taken from \cite{Reynolds}, in order to have a homogeneous measurement sample. They have been derived using the same type of fit for all spectra. Figure \ref{f1} displays $\tau_{\ion{O}{vii}}$ versus $\tau_{\ion{O}{viii}}$. There is a good correlation between the two parameters and $\tau_{\ion{O}{vii}}$ seems to be almost always greater than $\tau_{\ion{O}{viii}}$. In some cases, variability of the optical depths is observed (as for example \object{MCG-6-30-15}: Reynolds et al. \cite{Reynolds95} and \object{NGC 4051}: Guainazzi et al. \cite{Guainazzi}) and their position in this graph will change with respect to the Reynolds (\cite{Reynolds97}) values.

\indent In order to take into account the upper limits, we have calculated the mean value for the 20 objects in R97, using the Kaplan-Meier estimator with ASURV Rev 1.2 (Lavalley et al. \cite{Lavalley}), which implements the method presented in Feigelson $\&$ Nelson (\cite{Feigelson}). We obtain $\tau_{\ion{O}{vii}}$=0.33$\pm$0.07 and $\tau_{\ion{O}{viii}}$=0.20$\pm$0.07.

\subsection{Coronal lines}\label{sec:rc}

\begin{figure}
\resizebox{7.75cm}{!}{\includegraphics{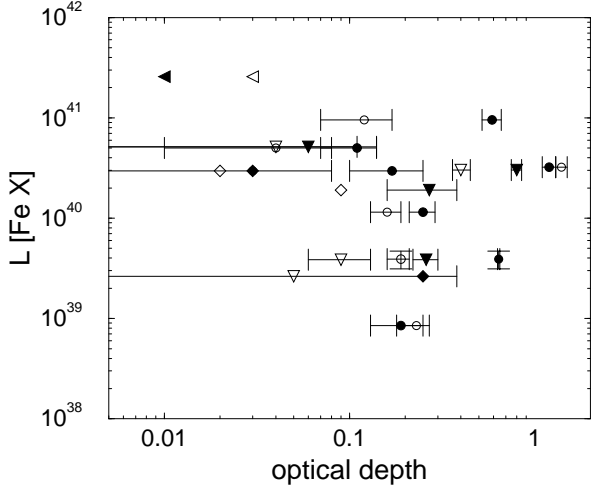}}
\caption{Observed dereddened [\ion{Fe}{x}] luminosity (erg.s$^{-1}$) versus $\tau_{\ion{O}{vii}}$ and $\tau_{\ion{O}{viii}}$. {\it Filled symbols}: $\tau_{\ion{O}{vii}}$ and {\it open symbols}: $\tau_{\ion{O}{viii}}$; {\it Triangle down}: upper limit for the Y-axis value (here the [\ion{Fe}{x}] luminosity), {\it triangle left}: upper limit for the X-axis value and {\it diamond}: upper limits for both X and Y axis values.}
\label{f3}
\end{figure}

\begin{figure}
\resizebox{7.65cm}{!}{\includegraphics{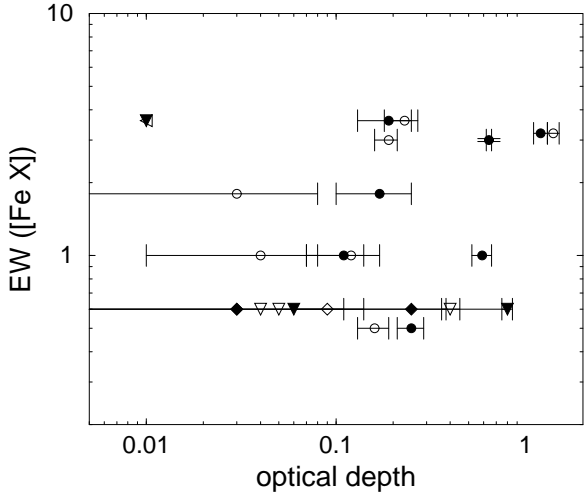}} 
\caption{Equivalent widths (in \AA) of [\ion{Fe}{x}] versus the optical depths of \ion{O}{vii} and \ion{O}{viii}. Same legend as Fig.~\ref{f3}.}
\label{f4}
\end{figure}

\begin{figure}
\resizebox{7.60cm}{!}{\includegraphics{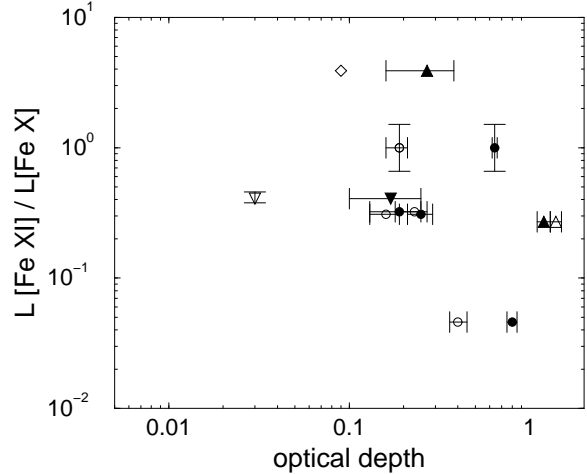}} 
\caption{Ratio of the observed dereddened luminosities of [\ion{Fe}{xi}] over [\ion{Fe}{x}] versus $\tau_{\ion{O}{vii}}$ and $\tau_{\ion{O}{viii}}$. Same legend as Fig.~\ref{f3} and {\it triangle up}: lower limit for the Y-axis value. [\ion{Fe}{xi}] data are from Reynolds et al. (\cite{Reynolds97}) (\object{MCG-6-30-15}), Penston et al. (\cite{Penston}) (\object{NGC 7469},  \object{Mrk 79}, \object{NGC 4051}, \object{NGC 5548}, \object{NGC 3516}, \object{Mrk 335}), and \cite{Erkens} (\object{Mrk 9}, \object{Mrk 704}, \object{Mrk 705}, \object{Mrk 1239}, \object{Akn 120}, \object{Mrk 699}, \object{Akn 564}).}
\label{f5}
\end{figure}

\indent Figure~\ref{f2} displays dereddened [\ion{Fe}{x}] 6375\AA\, luminosity versus soft X-ray luminosity integrated over 0.1-2.4 keV. The values have been compiled in the literature (see Ref. in Fig.\ref{f2} caption). 
There is a clear correlation between the two quantities, as also found by E97. This could be an indication that the high-ionization coronal lines, as the features of the WA, are formed either in a photoionized medium or in a medium which is partly photoionized and partly ionized by collisions. Figure~\ref{f3} displays the observed dereddened luminosity of [\ion{Fe}{x}] versus $\tau_{\ion{O}{vii}}$ and $\tau_{\ion{O}{viii}}$. Figure~\ref{f4} displays the [\ion{Fe}{x}] equivalent width (EW) versus oxygen optical depths. No correlations between these quantities are apparent. We have calculated (same method as defined above) a mean [\ion{Fe}{x}] equivalent width of 1.44$\pm$0.30$~$\AA. We take throughout this article EW[\ion{Fe}{x}]=1.5$~$\AA.\\
\indent For [\ion{Fe}{xi}] 7892\AA, we obtained a mean EW of about 4$~$\AA$~$ with only 3 EWs available for \object{MCG-6-30-15}, \object{NGC 3783} and \object{Mrk 1347} (Morris $\&$ Ward \cite{Morris}). 
Figure~\ref{f5} displays the luminosity ratio [\ion{Fe}{xi}]/[\ion{Fe}{x}] versus optical depths of \ion{O}{vii} and \ion{O}{viii}. For the reported objects, the [\ion{Fe}{x}] luminosity is greater than the [\ion{Fe}{xi}] luminosity, except for one object.\\
\indent For [\ion{Fe}{xiv}] 5303\AA, \cite{Erkens}, who have selected objects in which the presence of FHILs has already been reported in the literature, found that only 4 of their 15 objects ($\sim$27$\%$) required the presence of a significant [\ion{Fe}{xiv}] contribution. So this line is generally not detected in Seyfert 1s. The resolution and S/N of their spectra were not sufficient to separate the blend of [\ion{Fe}{xiv}] 5303\AA\,+\,[\ion{Ca}{v}] 5309\AA\ and no [\ion{Fe}{xiv}] EW values are quoted. Since only one spectrum has been plotted, we are thus unable to determine mean values for this coronal line. Nevertheless when [\ion{Fe}{xiv}] is present, its flux is significant ($>$25$\%$ of the [\ion{Ca}{v}] flux), then we will use EW[\ion{Fe}{xiv}]=3$~$\AA$~$ as a conservative upper limit. We will also use a value of 2 \AA$~$ to illustrate the sensitivity of this EW value on the physical parameters.\\

\indent Until now, only very little information concerning some infrared coronal lines has been published for Seyfert 1s: for \object{NGC 7469} ([\ion{Si}{x}]1.43$\mu$m and [\ion{S}{xi}]1.93$\mu$m: Thompson \cite{Thompson96}, [\ion{Fe}{xii}]2.20$\mu$m: Genzel et al. \cite{Genzel}) and for \object{NGC 3516} ([\ion{Ca}{viii}]2.32$\mu$m: Giannuzzo et al. \cite{Giannuzzo}). Mean EWs for these coronal lines cannot be determined. We will assume that they are smaller than 10$~$\AA, which is compatible with the data of \object{NGC 3516}.

\subsection{UV high-ionization resonance lines: \ion{O}{vi} 1034\AA$~$ and \ion{Ne}{viii} 774\AA}
Resonance lines correspond to allowed transitions to the ground level. Zheng et al. (\cite{Zheng97}) who have constructed a composite Radio-Quiet quasar spectrum with 101 quasars with z$>$0.33 gave: EW(\ion{Ne}{viii})=4\,\AA$~$. According to Figure 2 of Zheng et al. (\cite{Zheng95}), the EW(\ion{O}{vi}) for a sample of 24 Radio-Quiet Quasars is about 7\,\AA$~$ (the four atypical EWs $>>$20\,\AA$~$ being excluded).

\section{Calculations} \label{sec:Calculation}

\subsection{Previous models}

\indent The WA has been investigated by Netzer (\cite{Netzer93}, \cite{Netzer96}). In his photoionization code "ION", he considered not only the absorption properties but also the emission and reflection spectra, which were not included in previous computations. He showed that intense X-ray lines as well as a non negligible reflection continuum might be produced. So the spectral shape could be changed significantly with respect to the pure transmitted spectrum, especially around the absorption edges which are reduced. He computed the UV and soft X-ray line intensities for a large range of parameters and found that the strongest X-ray lines should have EWs of about 5--50\AA. Indeed, George et al. (\cite{George98}) showed that the introduction of the X-ray emission lines  significantly improves the fit of X-ray spectra. Another computation of the emission and reflection spectra has been performed by Krolik $\&$ Kriss (\cite{Krolik95}), in the absence of thermal equilibrium.\\
\indent The thermal stability of the WA has been discussed by Reynolds $\&$ Fabian (\cite{Reynolds Fabian}), using the photoionization code ``CLOUDY'' (Ferland \cite{Ferland91}). They showed the importance of the shape of the ionizing continuum. Also Krolik $\&$ Kriss (\cite{Krolik95}) have studied the thermal stability of the WA.\\

\indent Since coronal lines are associated with highly ionized ions, the creation of such ions requires a highly powerful process. Two main models could explain the emission of the coronal lines: a hot gas with collisional ionization, with T$>$10$^{6}$K (Oke $\&$ Sargent \cite{Oke}; Nussbaumer $\&$ Osterbrock \cite{Nussbaumer}) and a gas photoionized by a hard UV-X-ray continuum, with T$\sim$ a few 10$^{4}$K (Osterbrock \cite{Osterbrock}; Grandi \cite{Grandi}; Korista $\&$ Ferland \cite{Korista}, Oliva et al. \cite{Oliva}, Moorwood $\&$ Oliva \cite{Moorwood}). A third model involving photoionization plus shocks inside the NLR has also been proposed (Viegas-Aldrovandi $\&$ Contini \cite{Viegas-Aldrovandi}).\\ 
\indent With a sample of 15 Seyferts (including 11 Seyfert 1s), \cite{Erkens} found that the observed line strengths appear compatible with the predictions of a simple photoionized model calculated by Korista $\&$ Ferland (\cite{Korista}) and Spinoglio $\&$ Malkan (\cite{Spinoglio}). The former gave flux ratios of iron coronal lines for very low densities (n$_{H}\leq$10 cm$^{-3}$). For \object{PG1211+143}, Appenzeller $\&$ Wagner (\cite{Appenzeller}) observed a flux ratio of [\ion{Fe}{vii}] 6087\AA/[\ion{Fe}{x}] 6375\AA=0.8$\pm$0.2 compatible with their calculations. The latter made computations for the infrared fine-structure lines for higher densities (10$^{2}\leq$n$_{H}\leq$10$^{6}$ cm$^{-3}$), and for low ionization parameters (see~$\S$\ref{sec:our model} for a definition). Their predictions for [\ion{Mg}{viii}], [\ion{Si}{vii}], [\ion{Si}{ix}], [\ion{Si}{x}], are referred to the optical forbidden line [OIII] 5007\AA$~$ assuming both lines coming from the same medium whereas they are probably produced by two different media. Both articles (Korista $\&$ Ferland and Spinoglio $\&$ Malkan) used an old version of ``CLOUDY'' with inaccurate electronic collision strengths for iron ions (Mason \cite{Mason}): resonance effects near the threshold energy, which have a significant influence on the coronal line fluxes, are not included. These resonance effects are discussed in Dumont $\&$ Porquet (1998, hereafter referred as \cite{Dumont}). For example, the electronic collision rates published by Storey et al. (\cite{Storey}) and Pelan $\&$ Berrington (\cite{Pelan}) (as used in our code for [\ion{Fe}{xiv}] and [\ion{Fe}{x}] respectively) are much larger than the previous data, especially at low temperatures.

\subsection{Our model} \label{sec:our model}
\indent Our calculations are based on a new code IRIS, which computes detailed multi-wavelength spectra of photoionized and/or collisionally ionized gas, using as input the thermal and ionization structure computed by the photoionization code PEGAS (for more information about these two codes see \cite{Dumont}). It includes a large number of levels and splits the multiplets, thus providing an accurate spectrum from the soft X-rays to the infrared.\\
\indent All important atomic processes were taken into account: collisional electronic excitation and ionization, three-body recombination, photoionization, radiative and dielectronic recombination, excitation-autoionization and proton impact excitation. Charge transfer has not been implemented for the ground transitions of the coronal ions, since it is negligible in our calculations.\\
\indent For [\ion{Fe}{x}], the effective electronic collision strengths are taken from Pelan $\&$ Berrington (\cite{Pelan}) and the proton collision rates from Bely $\&$ Faucher (\cite{Bely}). For [\ion{Fe}{xi}] there are no published data available for ground transition effective collision strengths, taking into account the resonance effects, so we have computed them with an IDL subroutine of the CHIANTI database (Dere et al. \cite{Dere}). The proton collision rates for [\ion{Fe}{xi}] are from Landman (\cite{Landman}). For [\ion{Fe}{xiv}], the effective collision strengths are from Storey et al. (\cite{Storey}) and the proton collision rates for ground level transitions are from Heil et al. (\cite{Heil}).\\
\indent The element abundances are from Allen (\cite{Allen}).\\

\indent Our model assumes optically thin clouds in plane-parallel geometry with constant hydrogen density, in ionization equilibrium and surrounding a central source of radiation. The grid of parameters investigated here is:

\begin{enumerate}
\item Hydrogen density: $10^{8}\leq n_{H}\leq10^{12}$ cm$^{-3}$
\item Hydrogen column density: $10^{20}\leq N_{H}\leq 5\,10^{23}$ cm$^{-2}$
\item Ionization parameter: $2\leq \xi \leq4000$ erg\,cm\,s$^{-1}$
\end{enumerate}
with $\xi=\frac{L}{n_{H}~R^{2}}$
\noindent where $L$ is the bolometric luminosity (erg.s$^{-1}$) integrated over 0.1 eV to 100 keV and R is the distance from the ionizing radiation source at the illuminated face of the cloud (cm).\\
\indent The WA is seen in about 50$\%$ of Seyfert 1s, so we suppose that it is present in all these galaxies with a covering factor of 0.5.\\
\indent Finally, we assume that no dust near the BLR could survive so close to the ionizing central radiation.\\

\indent We also compute an hybrid model which consists in a photoionized gas out of thermal equilibrium (contrary to the pure photoionized gas). In this case the temperature is set constant at 10$^{6}$K.

\begin{figure}
\resizebox{7.75cm}{!}{\includegraphics{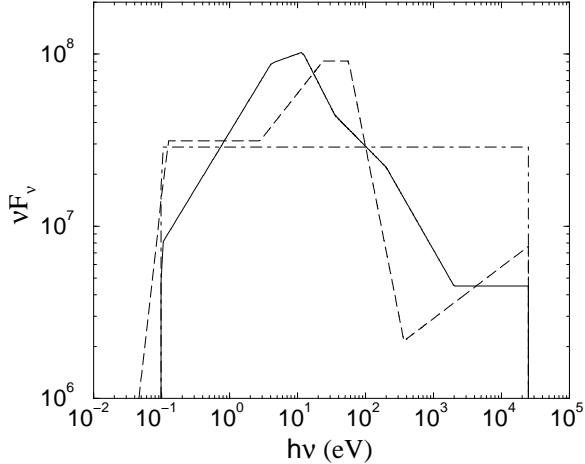}}
\caption{Shape of the two incident continua for Radio-Quiet quasars used in this paper, as well as a power law continuum (F$_{\nu}\propto\nu^{-1}$) used for comparison. All continua are normalized at the same $\xi$. {\it Solid line}: Laor et al. (\cite{Laor}) (``Laor continuum''), {\it dashed line}: Mathews $\&$ Ferland (\cite{Mathews}) but with a break at 10$\mu$m (``AGN continuum'') and {\it dot-dashed line}: simple power law.}
\label{f6}
\end{figure}

\indent We use two typical incident continua for Radio-Quiet AGN which are displayed on Figure~\ref{f6}. The first one is published by Laor et al. (\cite{Laor}) (``Laor continuum''), and the second one is used in ``CLOUDY'' (version 9004, Ferland et al. \cite{Ferland98}) and is similar to that published by Mathews $\&$ Ferland (\cite{Mathews}) but with a sub-millimeter break at 10 $\mu$m  (``AGN continuum''). The ``Laor continuum'' in the X-ray range is based on ROSAT PSPC observations from the Bright Quasar Survey, with high S/N spectra. The selection of the objects in the sample is independent of their X-ray properties to avoid any bias. They found that the soft X-ray flux at $\sim$0.2--1 keV in Mathews $\&$ Ferland (\cite{Mathews}) is significantly underestimated. \\
\indent The thermal stability of clouds irradiated by both continua will be discussed in \cite{Dumont}.\\
\indent Emission due to radiative recombination is shaped as an exponential decay ($\propto~$exp($\frac{-(h\nu-\chi)}{kT}$)) with a width of kT and either appears as a hump or fills the absorption hollow near the ionization threshold and could be observed as a blueshift of the edge. In the case of photoionized models, kT is small and the \ion{O}{vii} or \ion{O}{viii} edges appear generally smoothed, while in the hybrid models (photoionized gas out of thermal equilibrium with a constant temperature T=10$^{6}$K), the hollow can be partly filled. Therefore, our comparisons with observations take into account this emission and the current spectral resolution of the X-ray spectra (ASCA).\\
\indent The EWs are calculated relatively to the attenuated (transmitted plus emitted) central continuum. The reflected continuum is small and has very little influence on the value of the EWs.

\section{Results for the mean observed Seyfert 1 features} \label{sec:results} 
\begin{figure*}
\resizebox{8.20cm}{!}{\includegraphics{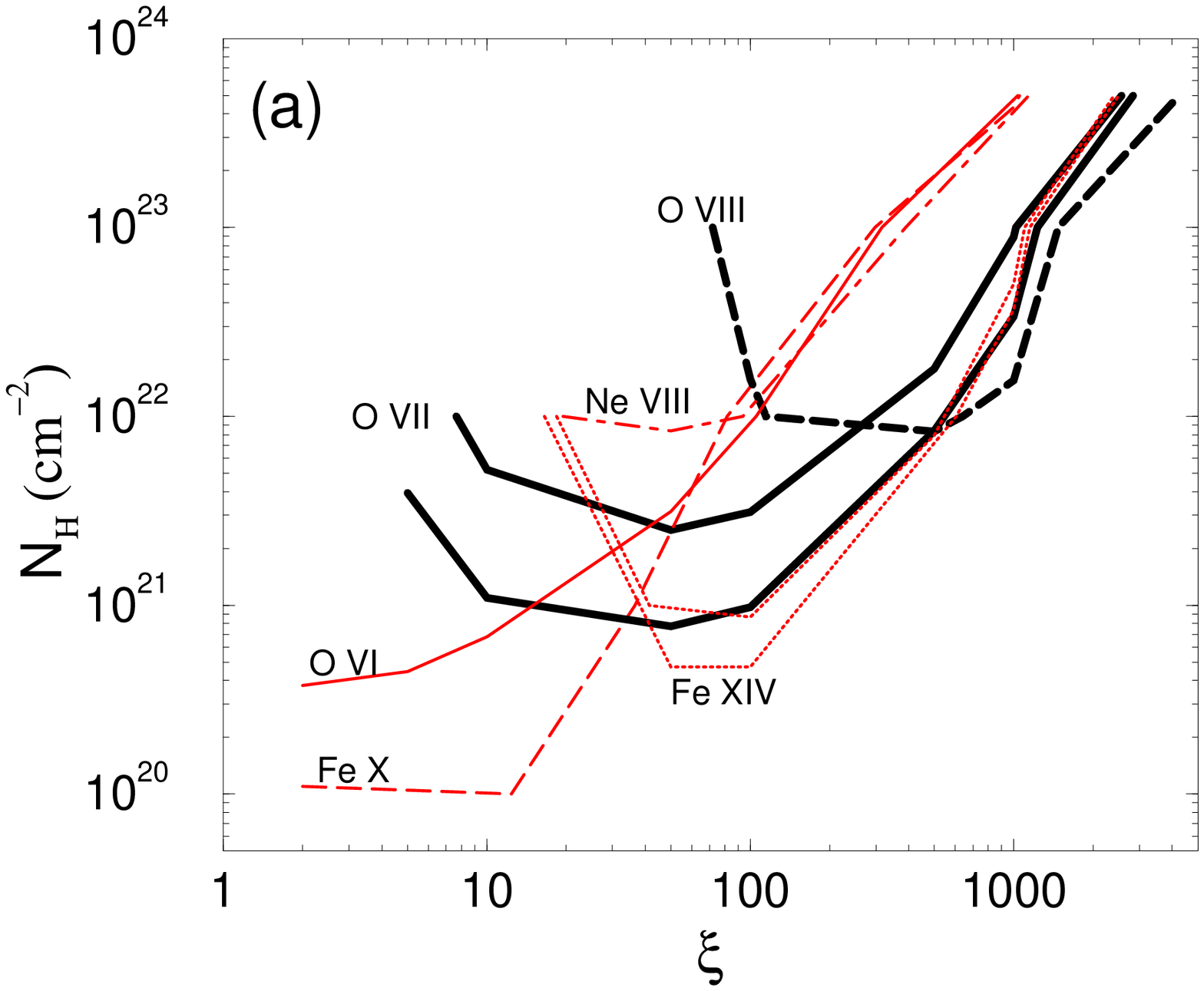}}\hspace{0.25cm}
\resizebox{8.20cm}{!}{\includegraphics{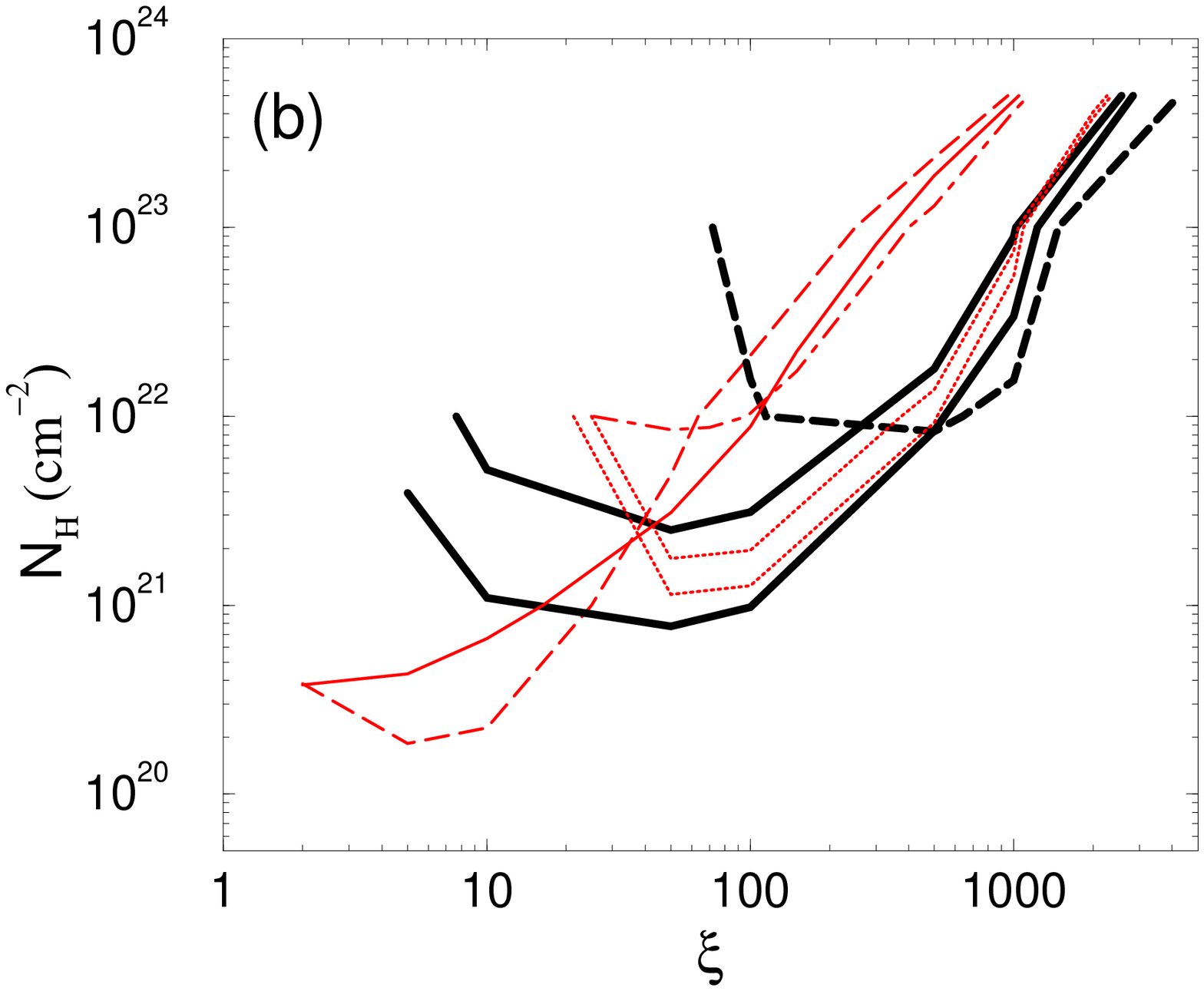}}

\resizebox{8.20cm}{!}{\includegraphics{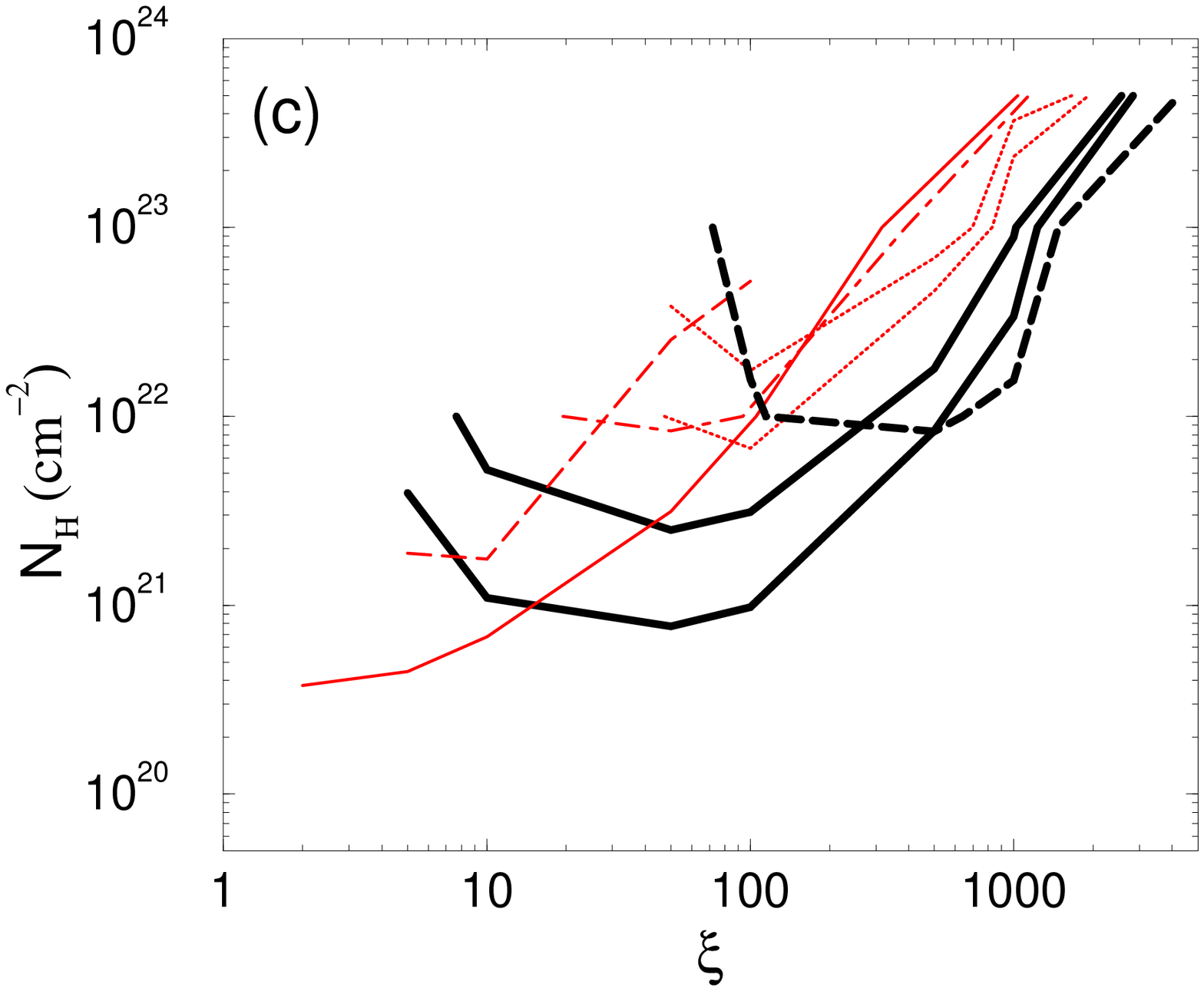}}\hspace{0.25cm}
\resizebox{8.20cm}{!}{\includegraphics{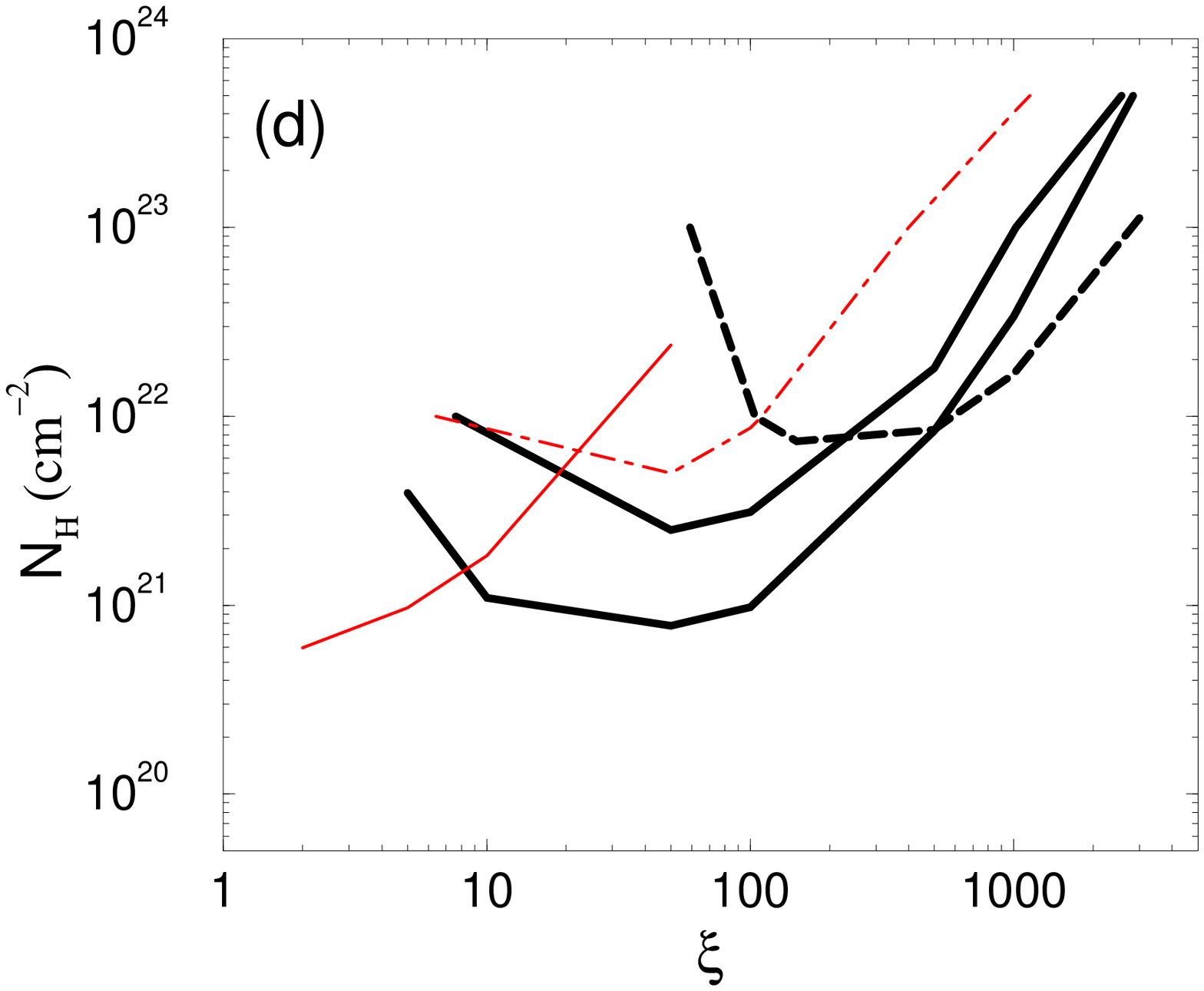}}
\caption{Isovalue curves in the plane ($\xi$,N$_{H}$) for the pure photoionized model with the incident ``Laor continuum'' for various hydrogen densities. (a) n$_{H}=10^{8}$ cm$^{-3}$, (b) n$_{H}=10^{9}$ cm$^{-3}$, (c) n$_{H}=10^{10}$ cm$^{-3}$ and (d) n$_{H}=10^{12}$ cm$^{-3}$. {\it Thick lower and upper solid lines}: $\tau_{\ion{O}{vii}}$=0.10 and 0.33 respectively, {\it thick long dashed line}: $\tau_{\ion{O}{viii}}$=0.20. {\it Thin long dashed line}: EW([\ion{Fe}{x}])=1.5$~$\AA, {\it Thin lower and upper dotted lines} EW([\ion{Fe}{xiv}])=2 and 3$~$\AA$~$ respectively, {\it thin dotted-dashed line}: EW(\ion{Ne}{viii})=4$~$\AA$~$ and {\it thin solid line}: EW(\ion{O}{vi})=7\AA. {\bf For the mean observed Seyfert 1 features, the regions above each thin curve are forbidden since producing too large EWs}.}
\label{f7}
\end{figure*}
\begin{center}
\begin{table*}
\caption{Results for the pure photoionized model with the incident ``Laor continuum'' for the mean observed Seyfert 1 features (cf Fig.~\ref{f7}).}   
\label{t1}
\begin{tabular}{|c|l|l|}
\hline
                   &                                                                              &  \\                                                
 $n_{H}$           & \multicolumn{1}{c|}{\ion{O}{vii} zone}                                       &  \multicolumn{1}{c|}{\ion{O}{viii} zone}  \\                     
 (cm$^{-3}$)                     & \multicolumn{1}{c|}{$\tau_{\ion{O}{vii}}$=0.33 ($\xi<$250)}    &\multicolumn{1}{c|}{$\tau_{\ion{O}{viii}}$=0.20 ($\xi>$250)}   \\
                   &                                                                              &  \\                                                
\hline
                   &                                                                             &    \\                                              
 10$^{8}$          & EW([\ion{Fe}{x}])$>>$1.5\AA$~$ or EW([\ion{Fe}{xiv}])$>>$3\AA               & $\bullet$ {\bf if} ($\xi\leq$600) {\bf then} EW([\ion{Fe}{xiv}])$>>$3\AA \\
                   & $\Longrightarrow$ this density is ruled out                                  & $\bullet$ {\bf else} no constraint  \\
                   &                                                                              &                                                     \\
\hline
                   &                                                                              &                                                 \\
 10$^{9}$          & idem as for n$_{H}$=10$^{8}$cm$^{-3}$                                        &  idem as for n$_{H}$=10$^{8}$cm$^{-3}$           \\
                   &                             &                                                  \\ 
\hline
                   &                                                                              &                                                   \\
 10$^{10}$         & $\bullet$ {\bf if} ($\xi\leq$40) {\bf then}  EW(\ion{O}{vi})$>>$7\AA                                  & no constraint        \\
                   & $\bullet$ {\bf else} no constraint                                           &                     \\
                   &                                                                              &                                                     \\
\hline
                   &                                                                              &                                                      \\
 10$^{12}$         & $\bullet$ {\bf if} ($\xi\leq$20) {\bf then} EW(\ion{O}{vi})$>>$7\AA                                  &  no constraint  \\ 
                   & $\bullet$ {\bf else} no constraint                                           &    \\
                   &                                                                              &                      \\
 \hline
\end{tabular}
\end{table*}
\end{center}
\begin{figure*}
\resizebox{8.20cm}{!}{\includegraphics{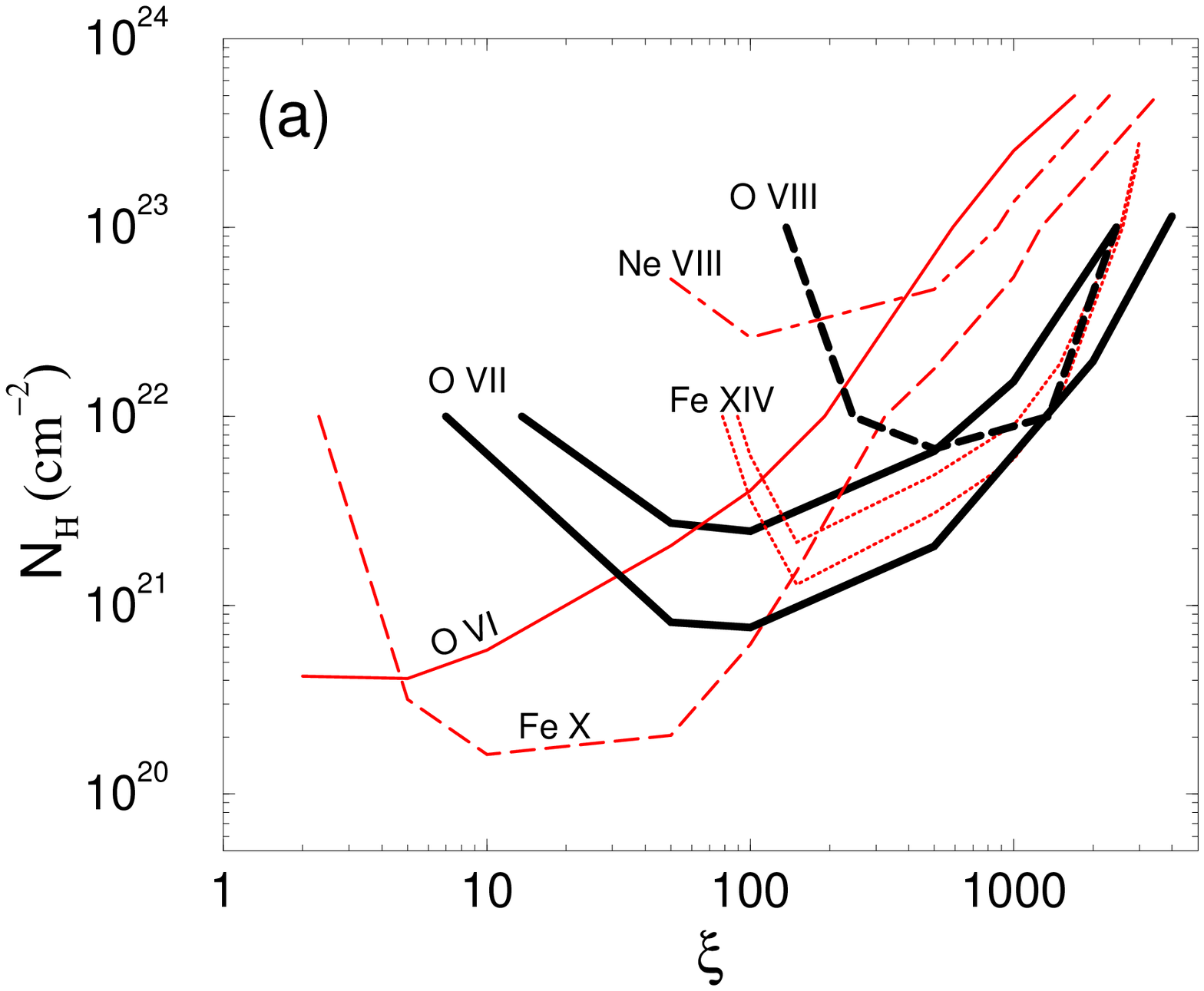}}\hspace{0.25cm}
\resizebox{8.20cm}{!}{\includegraphics{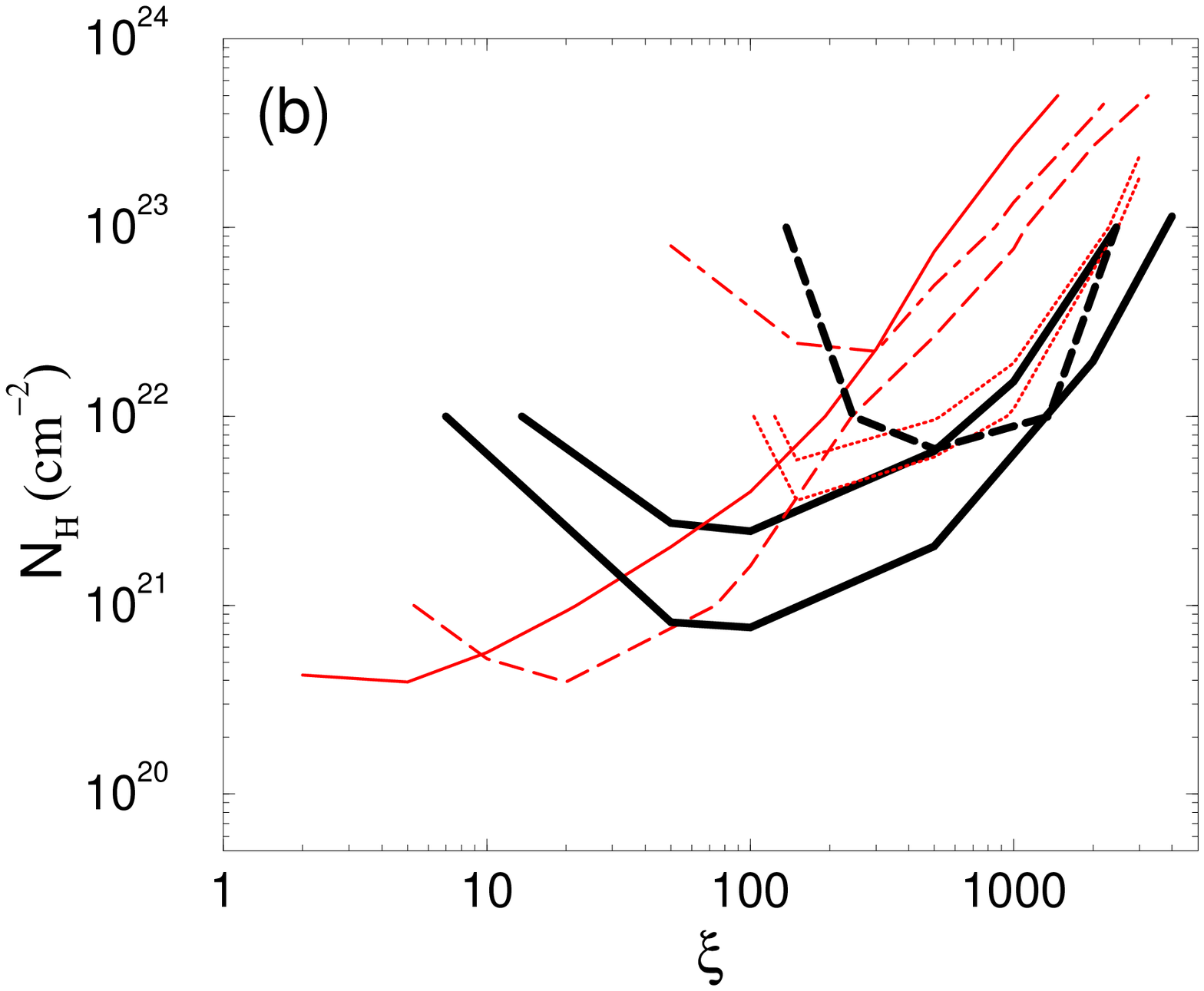}}

\resizebox{8.20cm}{!}{\includegraphics{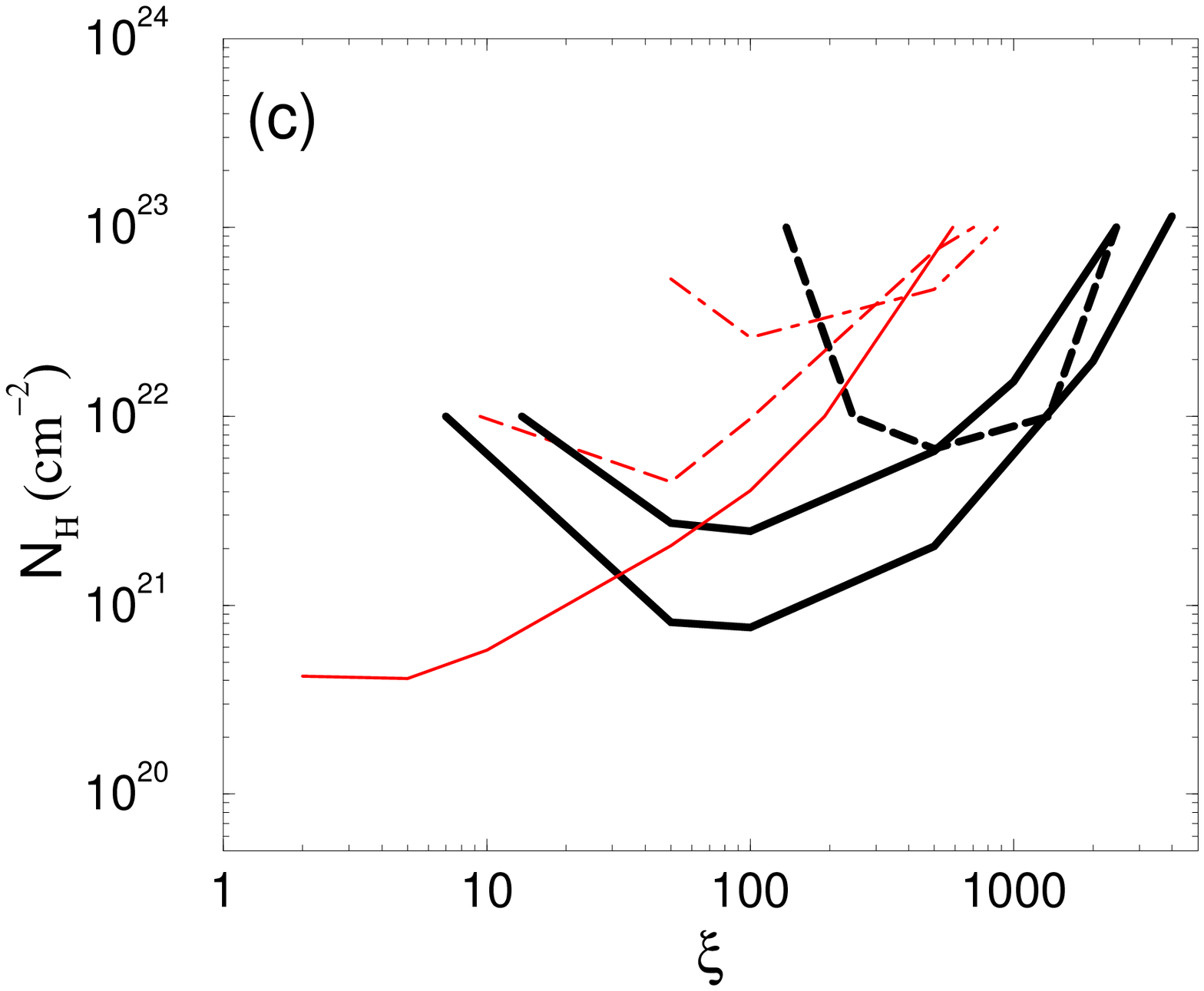}}\hspace{0.25cm}
\resizebox{8.20cm}{!}{\includegraphics{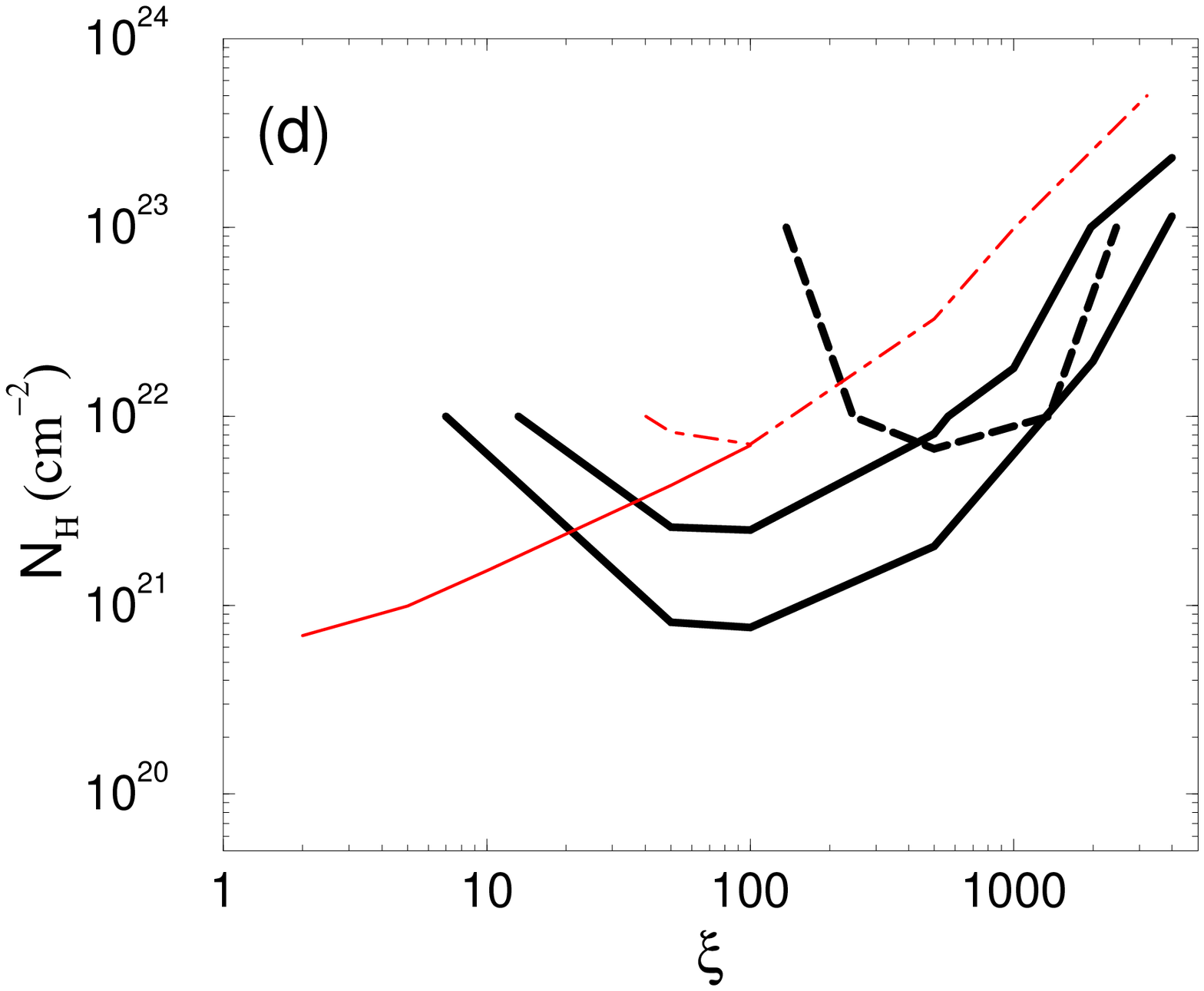}}
\caption{Same as Fig.~\ref{f7} for the pure photoionized model with the ``AGN continuum''.}
\label{f8}
\end{figure*}
\begin{figure*}
\resizebox{8.20cm}{!}{\includegraphics{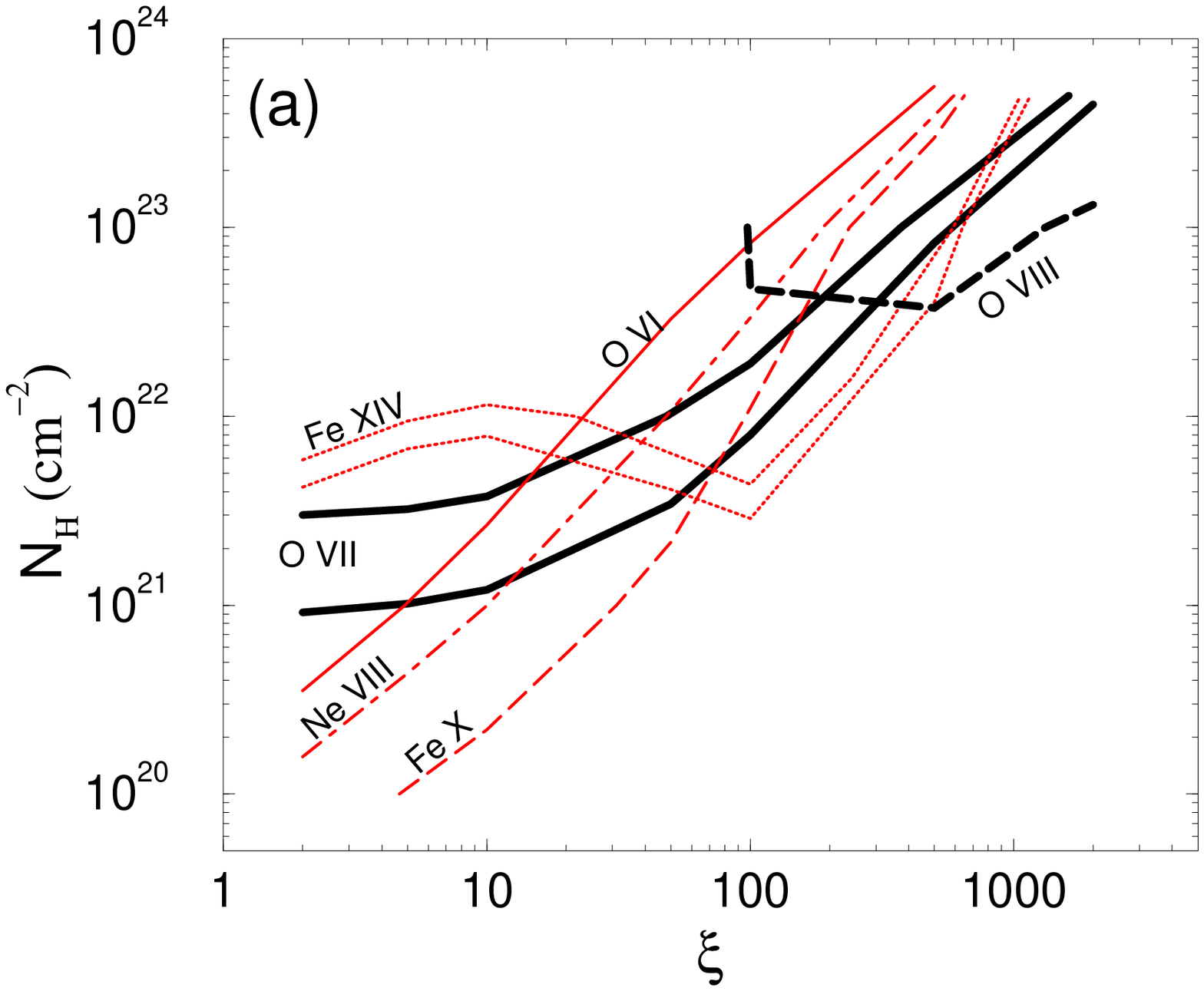}}\hspace{0.25cm}
\resizebox{8.20cm}{!}{\includegraphics{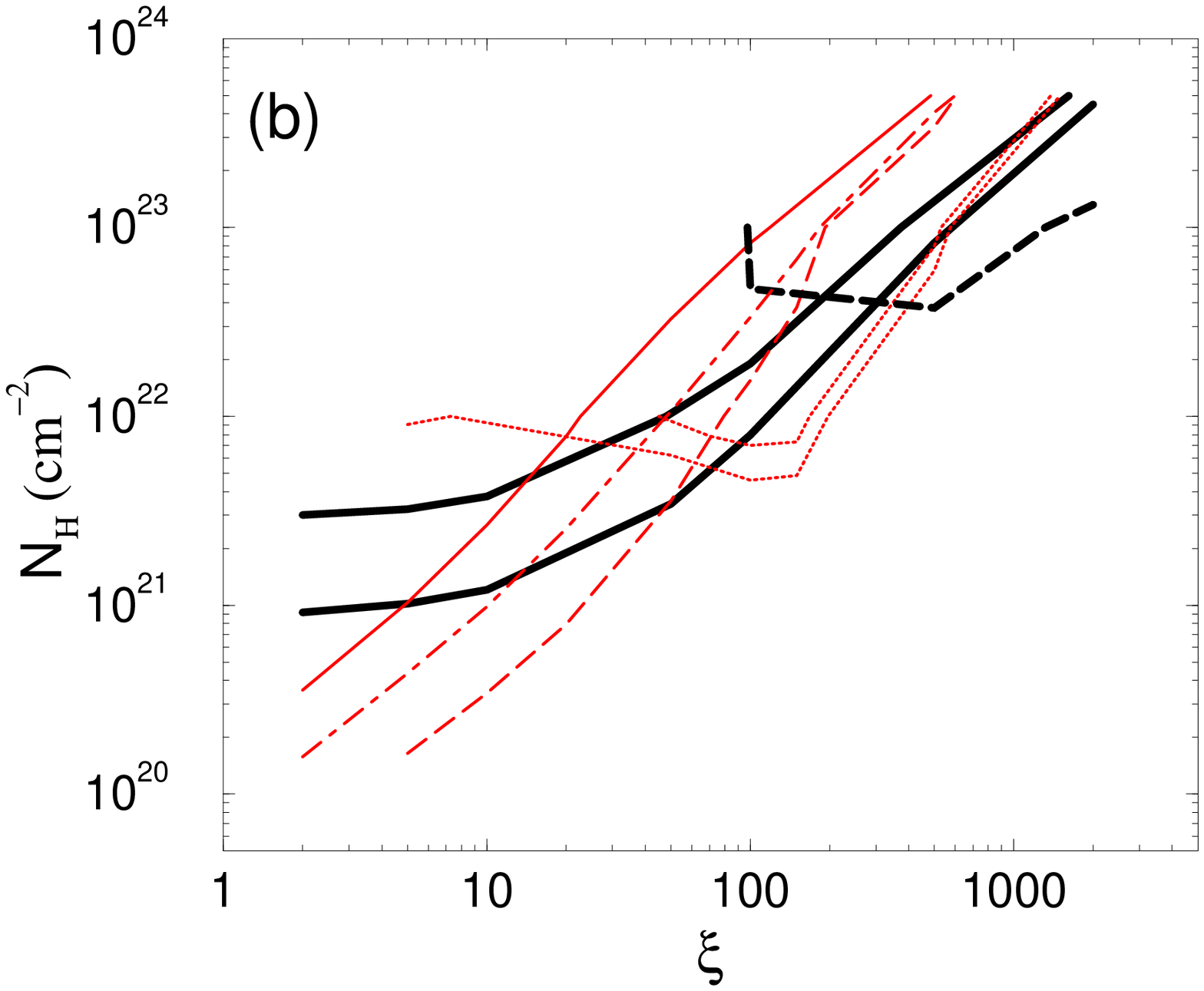}}

\resizebox{8.20cm}{!}{\includegraphics{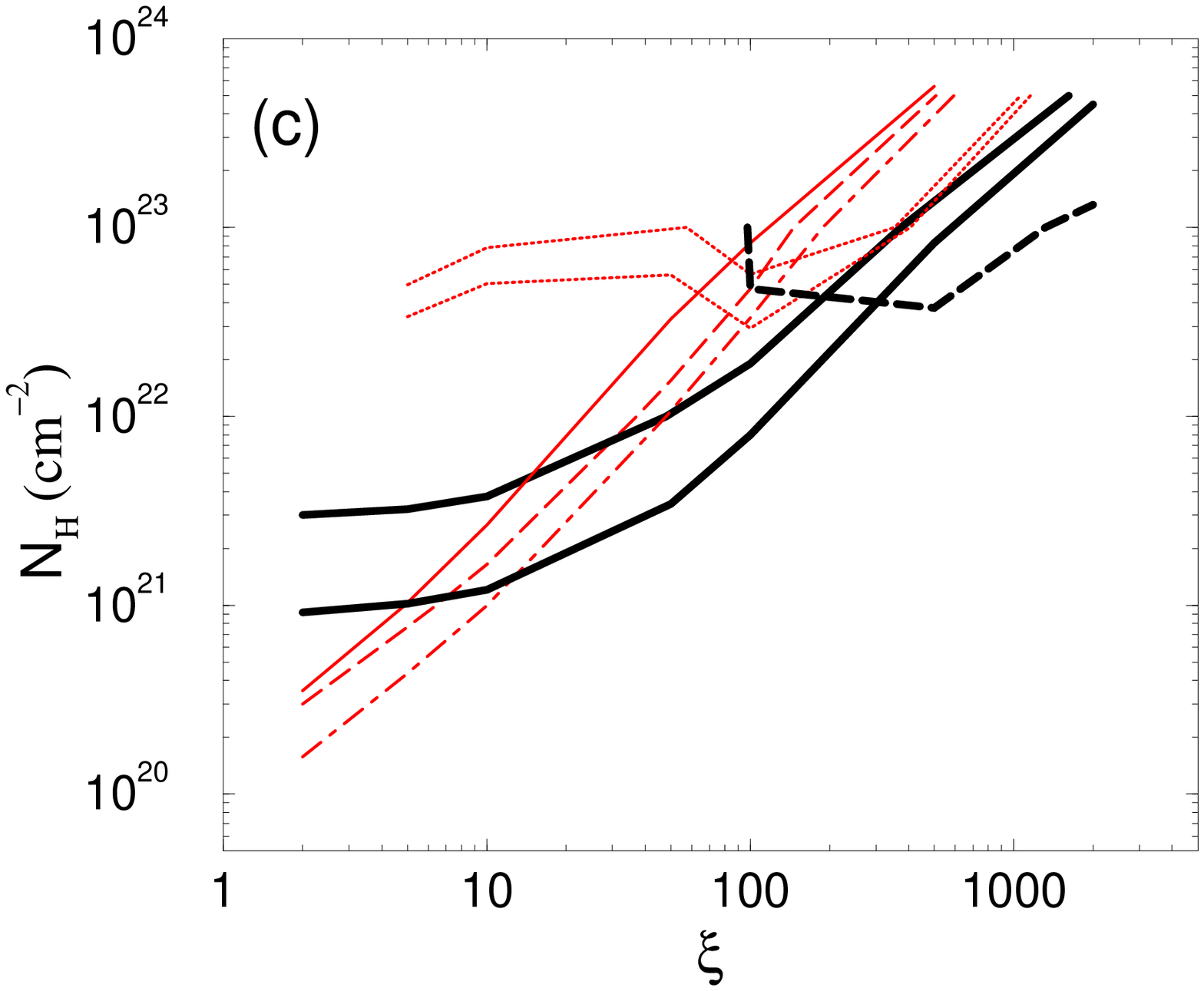}}\hspace{0.25cm}
\resizebox{8.20cm}{!}{\includegraphics{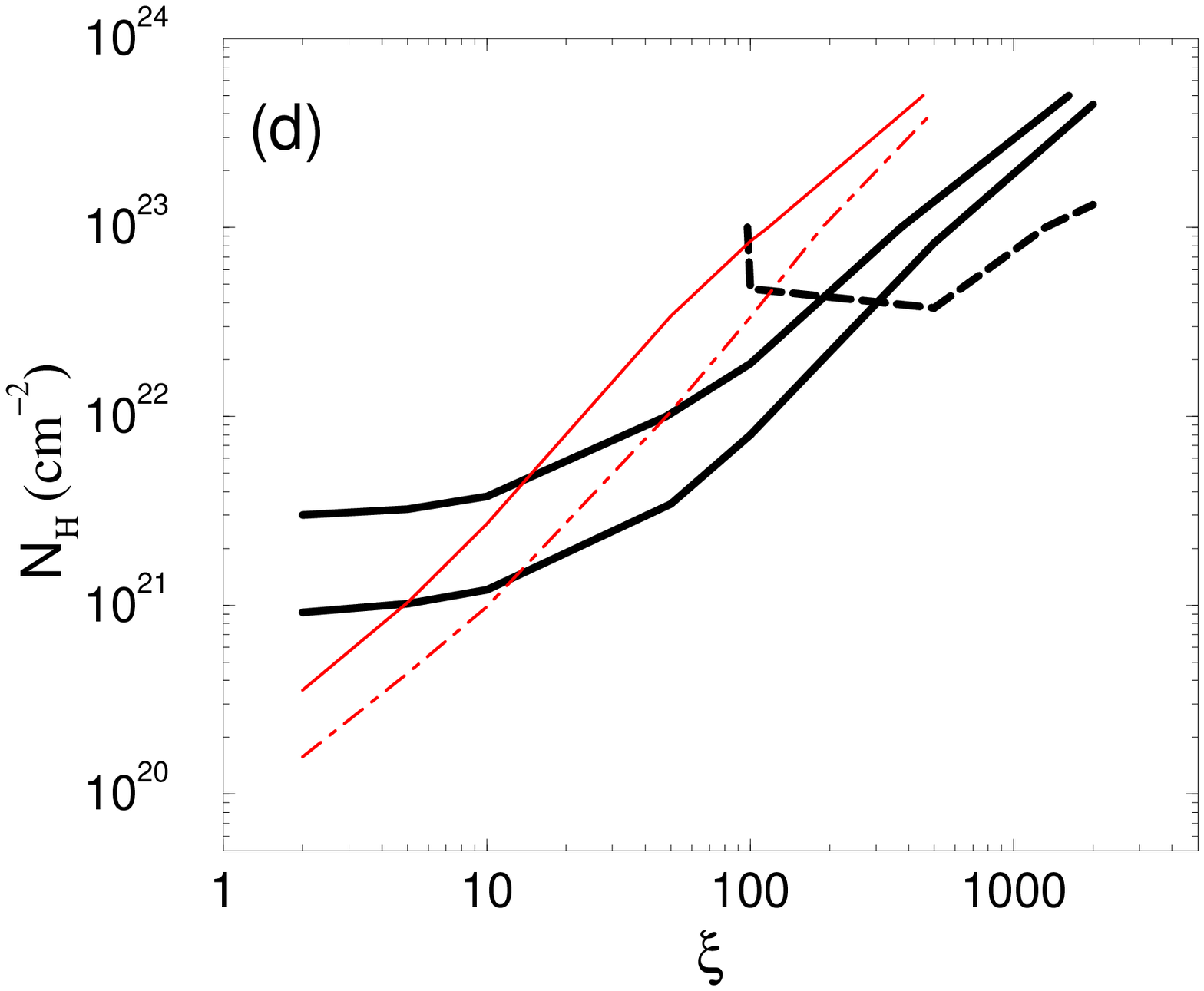}}
\caption{Same as Fig.~\ref{f7} for the hybrid model with the ``Laor continuum''.}
\label{f9}
\end{figure*}

\begin{figure*}
\resizebox{8.20cm}{!}{\includegraphics{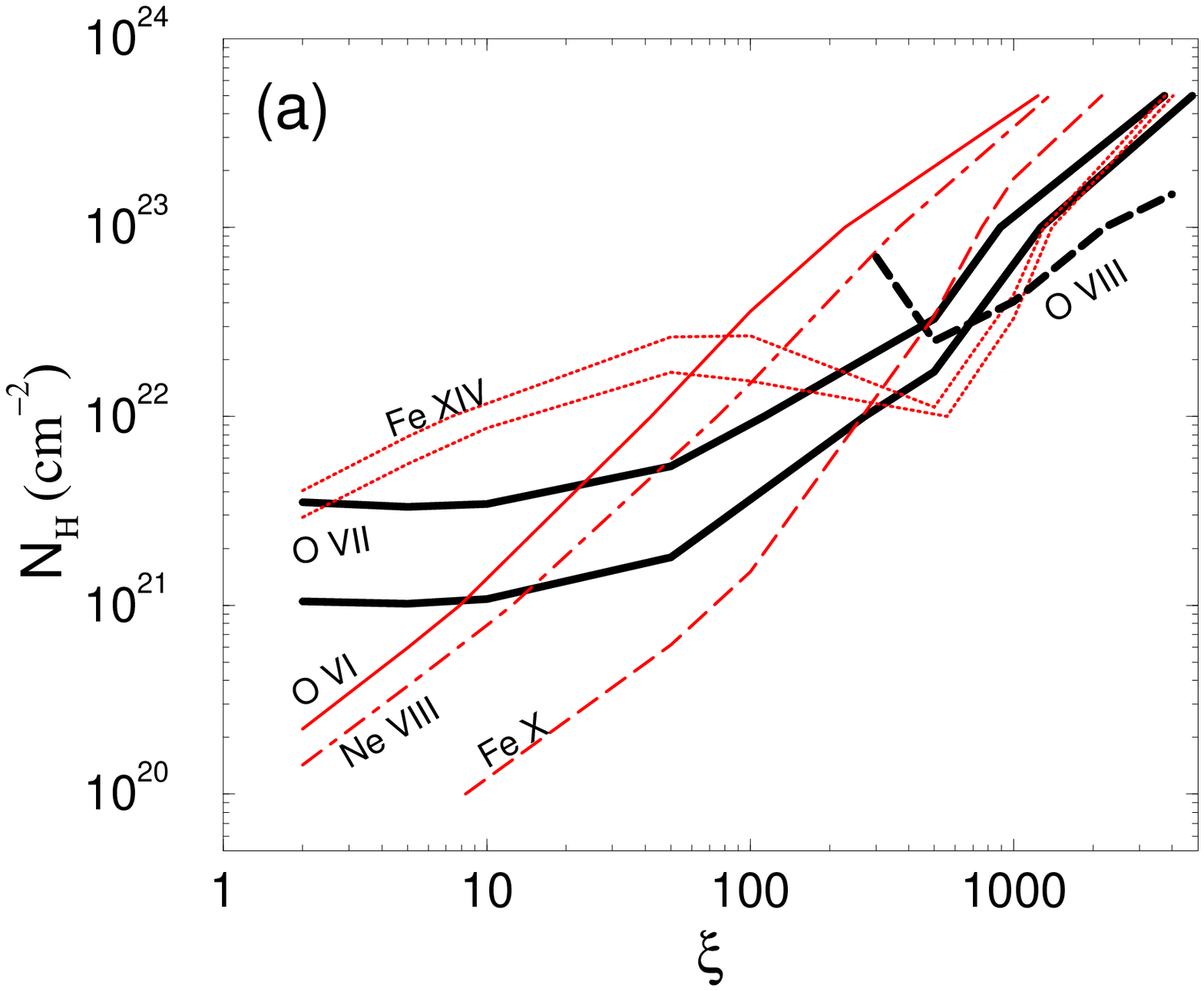}}\hspace{0.25cm}
\resizebox{8.20cm}{!}{\includegraphics{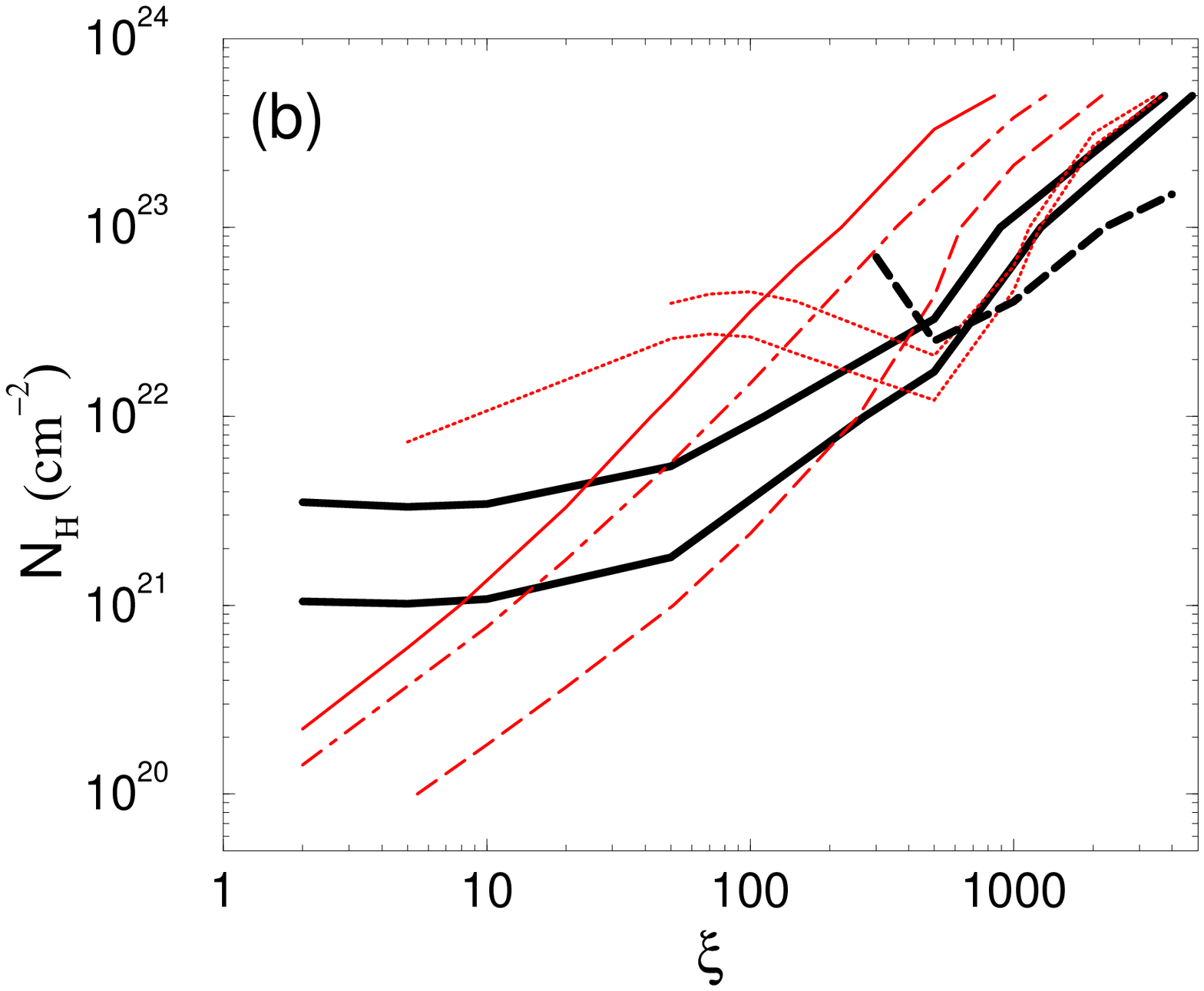}}

\resizebox{8.20cm}{!}{\includegraphics{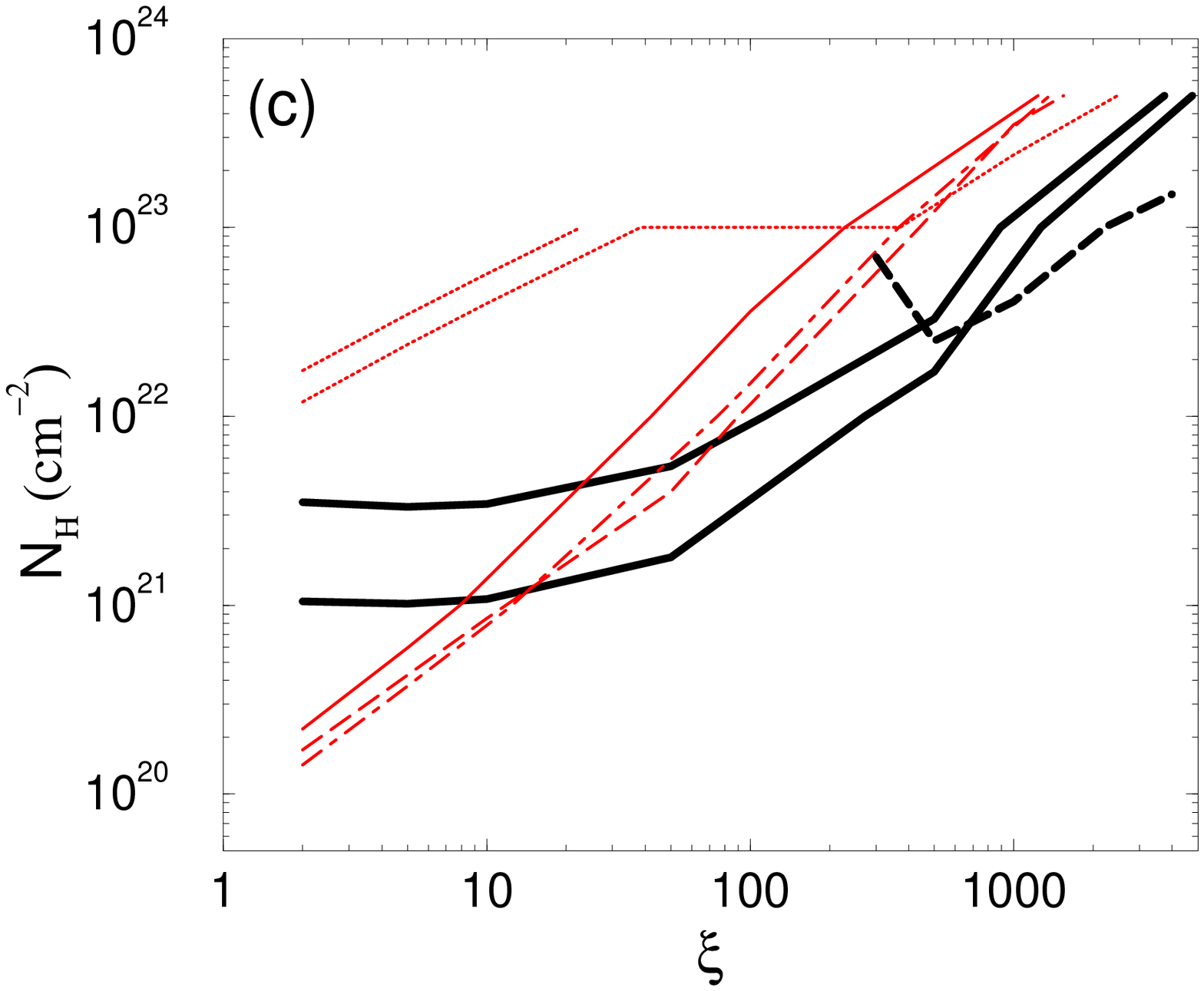}}\hspace{0.25cm}
\resizebox{8.20cm}{!}{\includegraphics{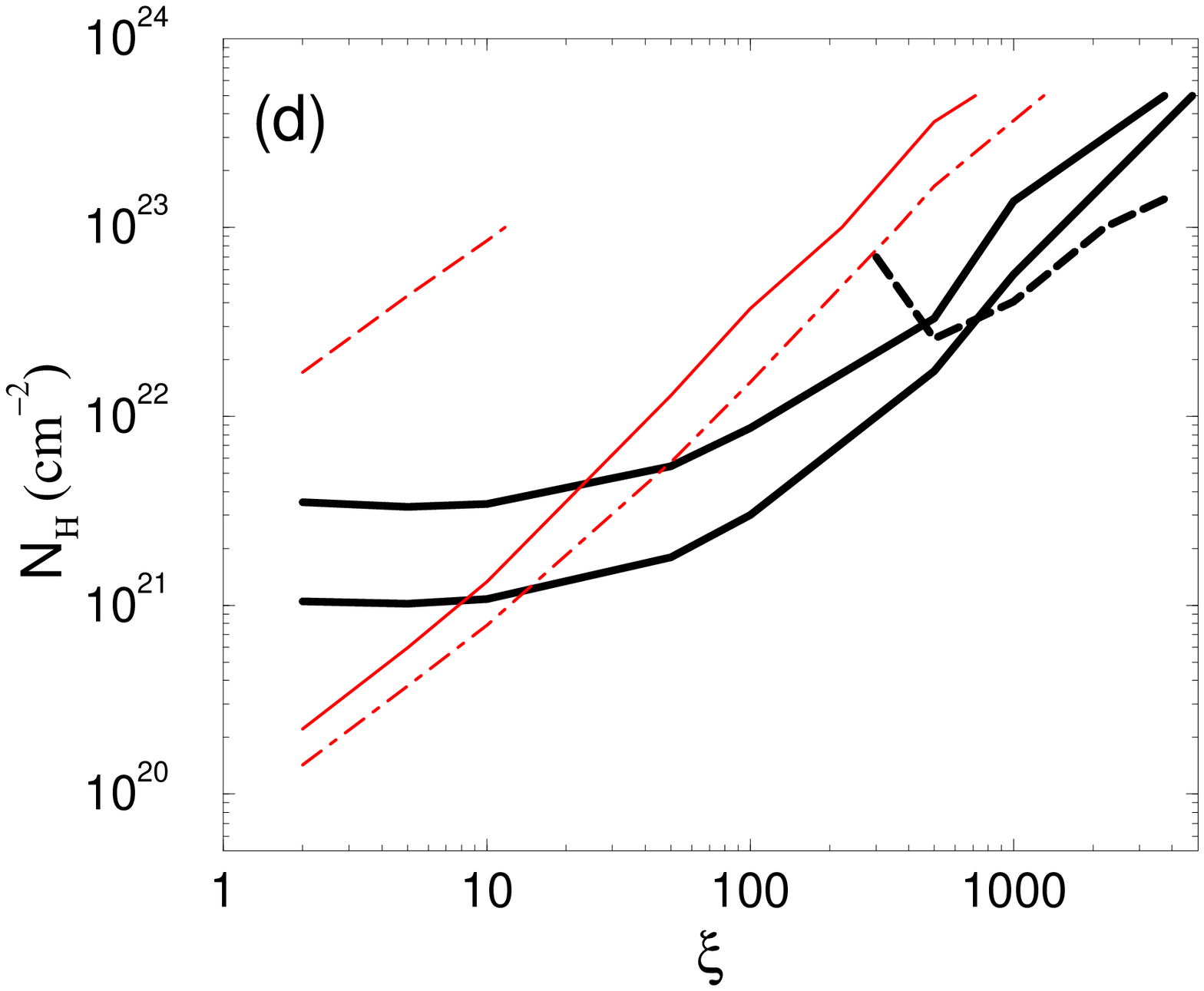}}
\caption{Same as Fig.~\ref{f7} for the hybrid model with the ``AGN continuum''.}
\label{f10}
\end{figure*}
\indent Figures \ref{f7}, \ref{f8}, \ref{f9} and \ref{f10} display, in the plane ($\xi$,N$_{H}$), isovalue curves corresponding to the {\bf mean} observed values of EWs and optical depths, in the case of the pure photoionized and hybrid models and for the ``Laor'' and ``AGN'' incident continua (n$_{H}$=10$^{8}$--10$^{9}$--10$^{10}$--10$^{12}$cm$^{-3}$). The thick long dashed, and the upper and lower solid curves represent $\tau_{\ion{O}{viii}}$=0.20 and $\tau_{\ion{O}{vii}}$=0.33 and 0.10, respectively. $\tau_{\ion{O}{vii}}$=0.33 and $\tau_{\ion{O}{viii}}$=0.20 are the mean values for Seyfert 1s (see $\S$\ref{sec:tau}). $\tau_{\ion{O}{vii}}$=0.10 is roughly the lower limit to detect the presence of the WA. EWs of coronal lines ([\ion{Fe}{x}] and [\ion{Fe}{xiv}]) and resonance lines (\ion{Ne}{viii} and \ion{O}{vi}) are also reported. They are mean values of EWs, except for [\ion{Fe}{xiv}] as explained in Sect.~\ref{sec:rc}. 
Grey thin long dashed, lower and upper dotted, dot-dashed and solid curves display isovalues of EW([\ion{Fe}{x}])=1.5$~$\AA, EW([\ion{Fe}{xiv}])=2 and 3$~$\AA, EW(\ion{Ne}{viii})=4$~$\AA\, and EW(\ion{O}{vi})=7$~$\AA$~$ respectively.\\
Isovalue curves for the IR coronal lines (upper limits taken at 10 \AA) and the [\ion{Fe}{xi}] coronal line (mean value of 4 \AA, based on three objects) are not displayed, since they do not constrain models more than the iron coronal lines [\ion{Fe}{x}] and [\ion{Fe}{xiv}] and than the resonance lines of \ion{Ne}{viii} and \ion{O}{vi}.\\

\indent For a given incident continuum shape and a given model (pure photoionized or hybrid), only a restricted range of $\xi$ and N$_{H}$ values is allowed to reproduce both $\tau_{\ion{O}{vii}}$=0.33 and $\tau_{\ion{O}{vii}}$=0.20. To avoid such a fine-tuning ($\xi\sim$250 and N$_{H}\sim$10$^{22}$ cm$^{-2}$ for the pure photoionized model with the ``Laor continuum''), a two-zone warm absorber is suggested, with the \ion{O}{vii} edge being formed in a region at lower $\xi$ and N$_{H}$ than the \ion{O}{viii} edge. Each zone responsible of a given edge has a negligible contribution to the other edge mainly formed in the second zone.\\  
\indent The regions of parameters ($\xi$,N$_{H}$) above the thin isovalue curves produce EWs for coronal and resonance lines greater than the mean observed values (except for [\ion{Fe}{xiv}] as explained above). Therefore, {\bf the region above each thin curve is forbidden}. 

\subsection{Pure photoionized models} 

\indent Figure~\ref{f7} shows the isovalue curves for the incident ``Laor continuum''. As an example, results for each density are displayed in Table~\ref{t1}. 
For $\tau_{\ion{O}{vii}}$=0.33, a high density (n$_{H}\geq$10$^{10}$cm$^{-3}$) is required, in order not to produce larger EWs of [\ion{Fe}{x}] or [\ion{Fe}{xiv}] than the mean observed ones. A very small region where $\tau_{\ion{O}{viii}}$=0.20 could correspond to n$_{H}$=10$^{8}$cm$^{-3}$ if N$_{H}\geq$10$^{22}$cm$^{-3}$ and $\xi\geq$600.

\indent Figure~\ref{f8} shows the isovalue curves for the case of the incident ``AGN continuum''. A one-zone model requires $\xi\sim$500 and N$_{H}\sim$7$\,$10$^{21}$ cm$^{-2}$. For  $\tau_{\ion{O}{vii}}$=0.33, a high density is required (n$_{H}>$10$^{9}$cm$^{-3}$). For $\tau_{\ion{O}{viii}}$=0.20, n$_{H}$ values as low as 10$^{9}$ cm$^{-3}$ are allowed.

\subsection{Results for the hybrid models}

\indent The pure photoionized model might be too simple to represent the WA, e.g. it could be in non-equilibrium or collisionally photoionized. So we have computed a grid for an hybrid model consisting in a photoionized gas out of thermal equilibrium, the temperature being taken constant at $T_{e}=10^{6}~$K. This temperature corresponds approximatively to the maximum of the ionic abundance of \ion{Fe}{x} and \ion{Fe}{xi} ions.\\
\indent Figure~\ref{f9} shows the isovalue curves for the incident ``Laor continuum''. Both edges could be produced in the same region for $\xi \sim$200 and N$_{H}\sim$5$\,$10$^{22}$ cm$^{-2}$. 
For $\tau_{\ion{O}{vii}}$=0.33, in order to have a non negligible allowed region, n$_{H}>$10$^{10}$cm$^{-3}$ is required with 50$<\xi<$200 and 10$^{22}<$N$_{H}<$5$\,$10$^{22}$ cm$^{-2}$. $\tau_{\ion{O}{viii}}$=0.20 could be accounted for by densities as low as n$_{H}\sim$10$^{8}$cm$^{-3}$ if N$_{H}\geq$4\,10$^{22}$cm$^{-2}$ and $\xi\geq$500.\\
\indent Figure~\ref{f10} displays the corresponding isovalue curves for the incident ``AGN continuum''. The one-zone model requires $\xi\sim$500 and N$_{H}\sim$3.5$\,$10$^{22}$ cm$^{-2}$. For $\tau_{\ion{O}{vii}}$=0.33, a high density is required (n$_{H}\geq$10$^{10}$ cm$^{-3}$) as in the previous cases. For $\tau_{\ion{O}{viii}}$=0.20, high ionization parameters ($\xi>$100) are required for low density values (n$_{H}\sim$10$^{8}$ cm$^{-3}$).

\subsection{Conclusions for the pure photoionized and for the hybrid models}

\indent The confrontation of the regions of parameters ($\xi$,N$_{H}$) allowed by the EWs with those producing $\tau_{\ion{O}{vii}}$ and $\tau_{\ion{O}{viii}}$ separately, strongly constrains the hydrogen density of the WA. For n$_{H}\leq$10$^{10}$ cm$^{-3}$ the physical parameters are mainly constrained by the coronal lines due to their weak observed EWs and their high critical densities. On the contrary, at higher densities constraints are given by the resonance lines.\\
\indent The isovalue curves between the models with the ``Laor'' and ``AGN'' continua are shifted by a factor of about 2 in $\xi$ due to continuum shape differences. In the same way, similar values of the optical depths for the hybrid case are obtained for a $\xi$ value five times smaller than for the pure photoionized case. Notice that for the hybrid model, the \ion{Ne}{viii} line is enhanced.\\
\indent A one-zone model which would be responsible for all features considered here is ruled out by all considered models.\\
\indent For both pure photoionized and hybrid models a high density n$_{H}\geq$10$^{10}$cm$^{-3}$ for the WA is required for 
$\tau_{\ion{O}{vii}}$=0.33, in order to explain the mean observed coronal lines and resonance lines of Seyfert 1 galaxies. On the contrary, $\tau_{\ion{O}{viii}}$=0.20 could be obtained with n$_{H}$ as low as 10$^{8}$cm$^{-3}$ but for a smaller range of parameters corresponding to high $\xi$ and N$_{H}$ values.

\section{Example of a particular case: MCG-6-30-15} \label{sec:mcg6-30-15}

\begin{figure}
\resizebox{8.20cm}{!}{\includegraphics{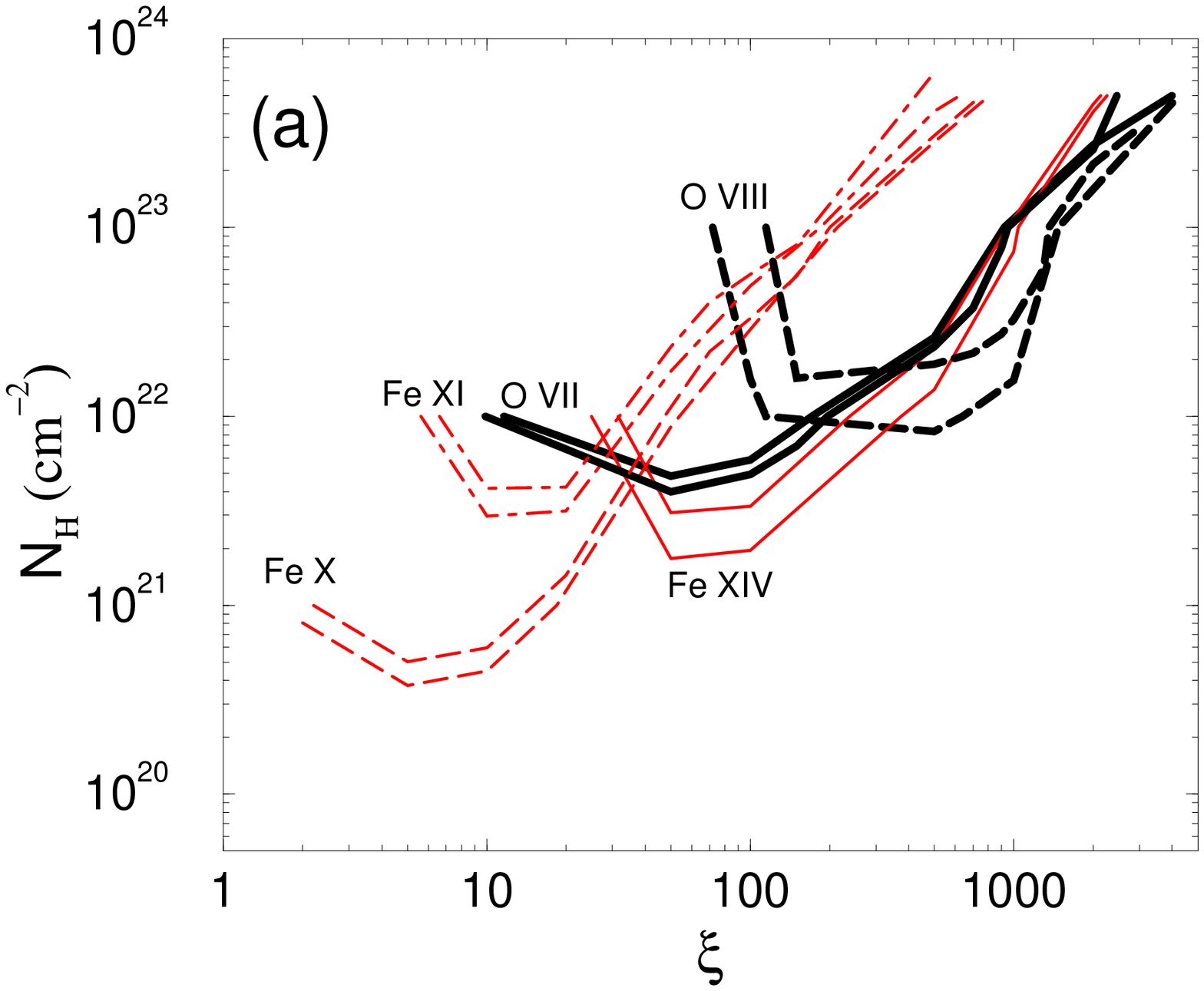}}
\resizebox{8.20cm}{!}{\includegraphics{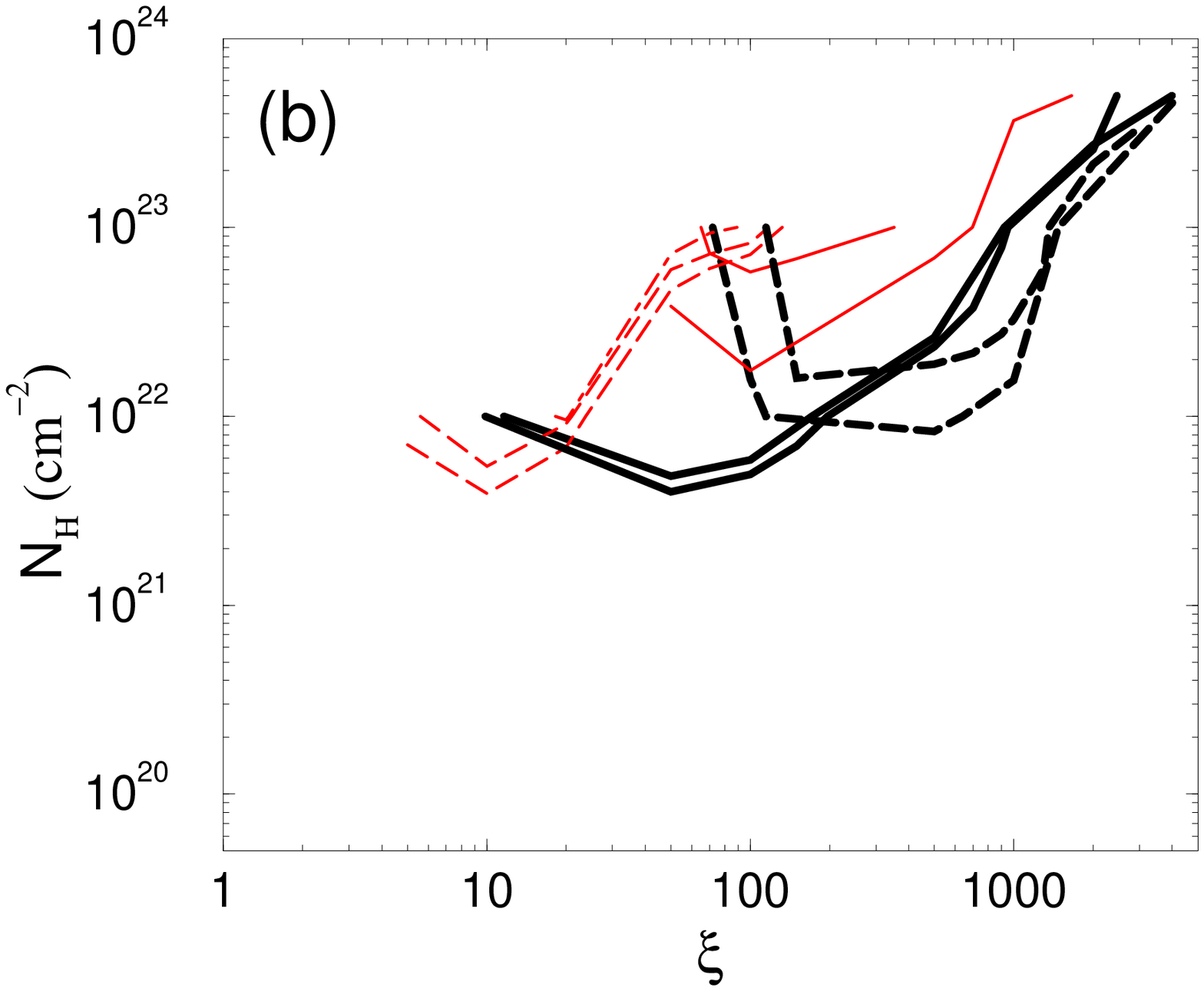}}
\caption{Isovalue curves for MCG-6-30-15 for the pure photoionized model with the ``Laor continuum'' for two different hydrogen densities: (a) n$_{H}$=10$^{9}$ cm$^{-3}$ and (b) n$_{H}$=10$^{10}$ cm$^{-3}$. {\it Thick lower and upper solid lines}: $\tau_{\ion{O}{vii}}$=0.53 and 0.63 respectively, {\it thick lower and upper dashed lines}: $\tau_{\ion{O}{viii}}$=0.19 and 0.44, {\it thin lower and upper long dashed lines}: EW([\ion{Fe}{x}])=3 and 4$~$\AA, {\it thin lower and upper dot-dashed lines}: EW([\ion{Fe}{xi}])=3 and 4$~$\AA, {\it thin lower and upper solid lines} EW([\ion{Fe}{xiv}])=3 and 5$~$\AA$~$ respectively.}
\label{f11}
\end{figure}

\begin{figure}
\resizebox{8.20cm}{!}{\includegraphics{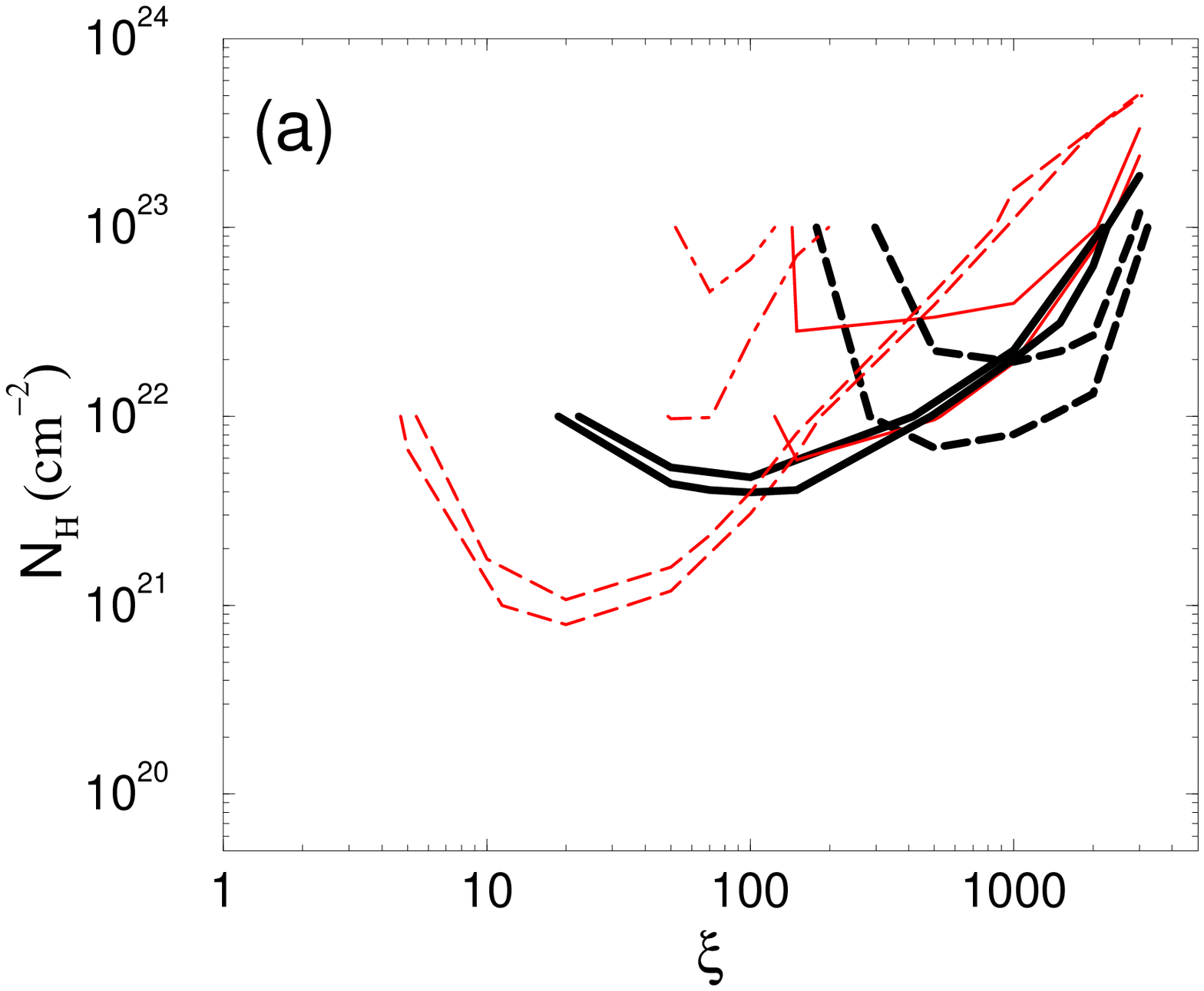}}
\resizebox{8.20cm}{!}{\includegraphics{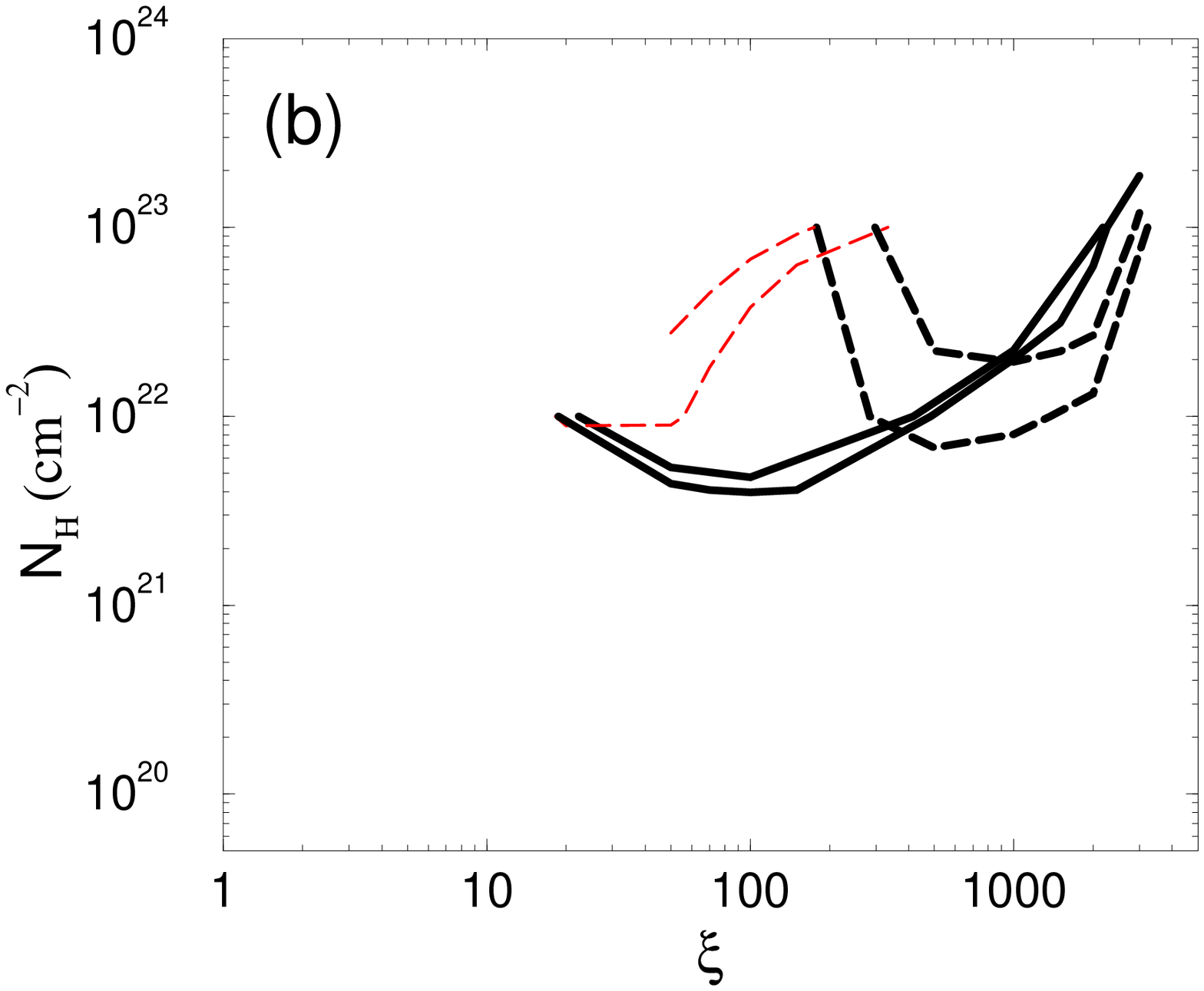}}
\caption{Same as Fig.~\ref{f11} for the pure photoionized models with the ``AGN continuum''.}
\label{f12}
\end{figure}

\begin{figure}
\resizebox{8.20cm}{!}{\includegraphics{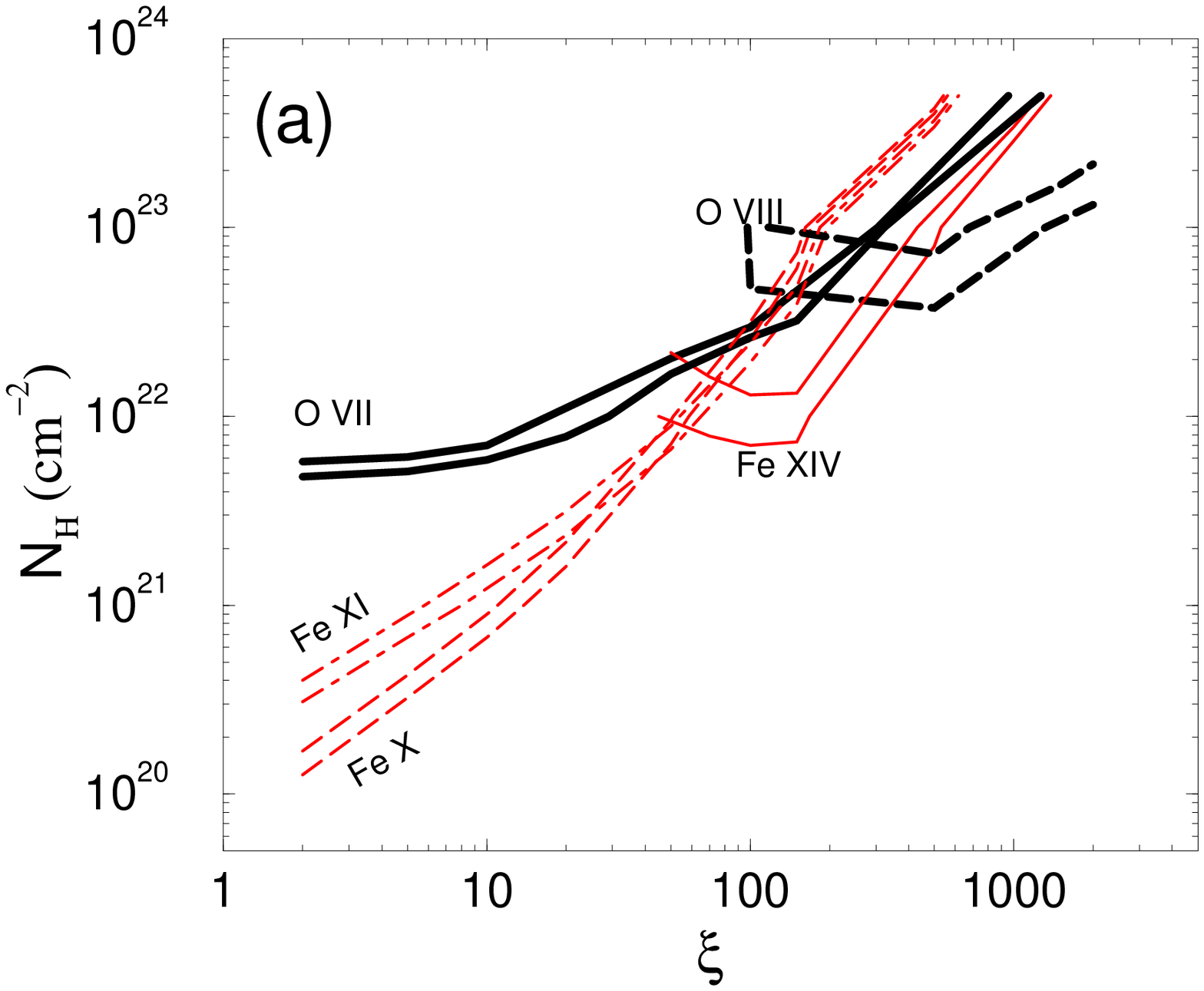}}
\resizebox{8.20cm}{!}{\includegraphics{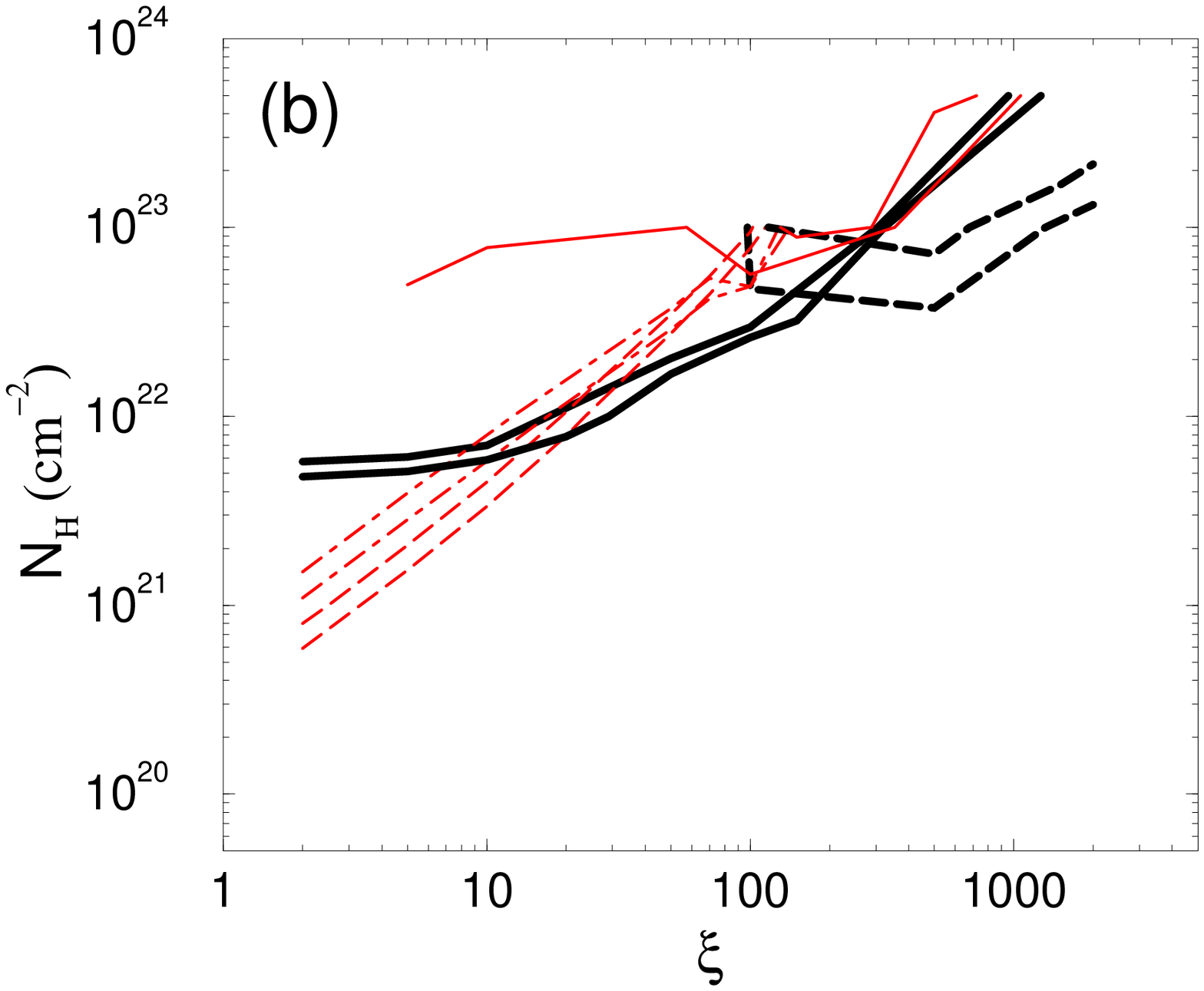}}
\caption{Same as Fig.~\ref{f11} for the hybrid models with the ``Laor continuum''.}
\label{f13}
\end{figure}

\begin{figure}
\resizebox{8.20cm}{!}{\includegraphics{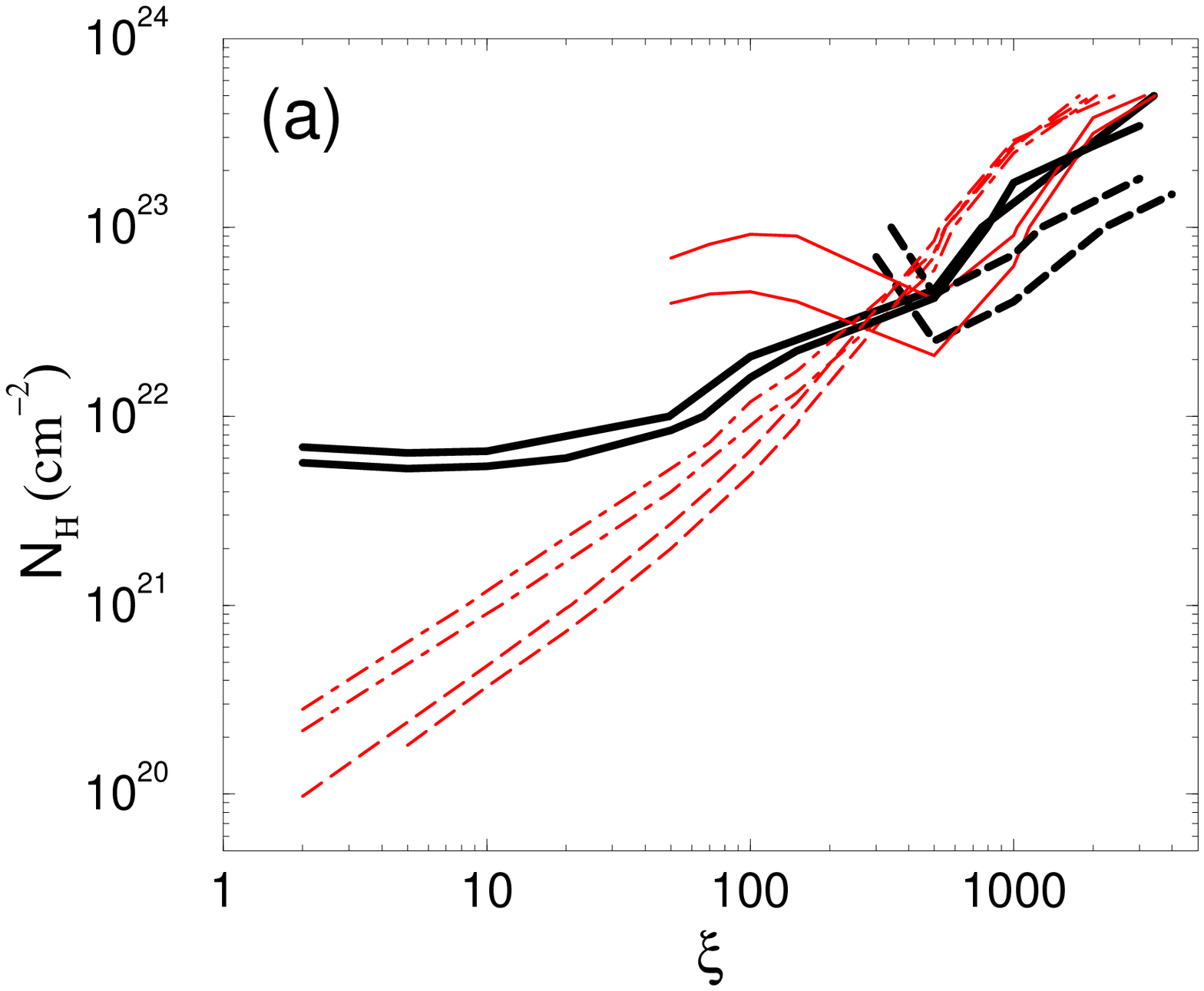}}
\resizebox{8.20cm}{!}{\includegraphics{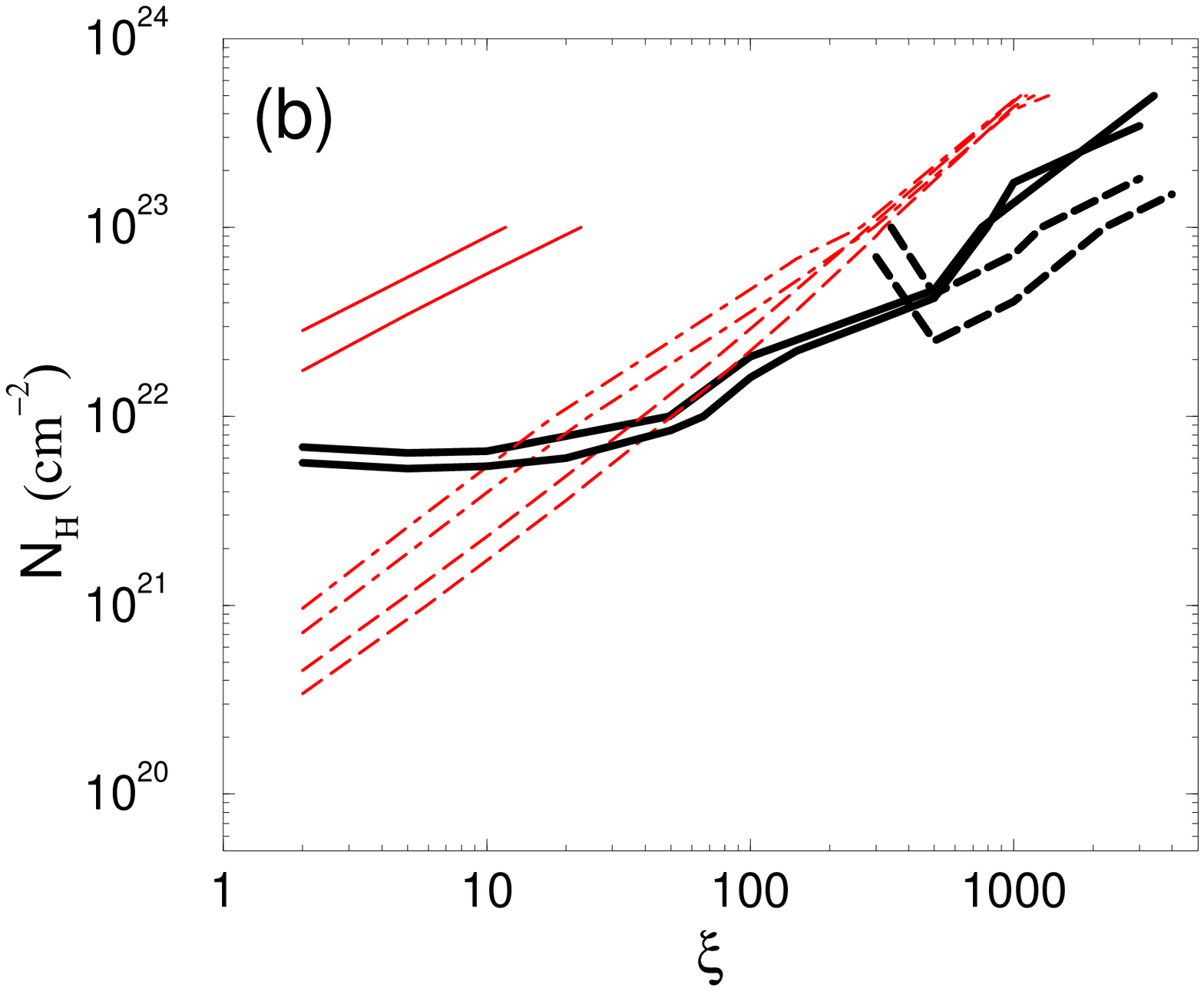}}
\caption{Same as Fig.~\ref{f11} for the hybrid models with the ``AGN continuum''.}
\label{f14}
\end{figure}

\indent It is interesting to apply our computations to the Seyfert 1 galaxy \object{MCG-6-30-15}, for which several emission lines have been measured, as well as the optical depths of the \ion{O}{vii} and \ion{O}{viii} edges.\\ 
\indent With ASCA, Reynolds et al. (\cite{Reynolds95}) observed in this object that the optical depth at the \ion{O}{viii} edge ($\tau_{\ion{O}{viii}}$) responds to continuum changes with a characteristic time-scale of about $10^{4}$s, whereas the optical depth of \ion{O}{vii} ($\tau_{\ion{O}{vii}}$) seems to be almost constant during all the observations. Otani et al. (\cite{Otani}) made the hypothesis that the WA could consist of two different components: the inner absorber which has a high ionization, responsible in large part of the \ion{O}{viii} edge and located in or just outside the BLR (R$<$10$^{17}$ cm), and the outer absorber mainly responsible of the \ion{O}{vii} edge, less ionized, and located near the molecular torus (R$>$1 pc). With BeppoSax data, Orr et al. (\cite{Orr}) confirmed that $\tau_{\ion{O}{vii}}$ did not change significantly during the exposure time ($\sim$400 ks), contrary to $\tau_{\ion{O}{viii}}$. But they did not find any simple correlation between $\tau_{\ion{O}{viii}}$ and the continuum luminosity, although the variations of the continuum emission and those of the WA (edges) have a similar time-scale ($<\,2.10^{4}$s). The ASCA mean spectrum obtained in July 1994 (Otani et al. \cite{Otani}) and the BeppoSax spectrum (Orr et al. \cite{Orr}) give optical depth values consistent with those derived from the 1993 ASCA data (Reynolds et al. \cite{Reynolds95}). In the following we use the two sets of values derived by Reynolds et al.: $\tau_{\ion{O}{vii}}$=0.53$\pm$0.04, 0.63$\pm$0.05 and $\tau_{\ion{O}{viii}}$=0.19$\pm$0.03, 0.44$\pm$0.04, respectively for the July and August 1993 datasets.\\
\indent Data for the EWs of iron coronal lines are taken from Reynolds et al. ({\cite{Reynolds97}). 
They gave EW([\ion{Fe}{x}])=3.0$\pm$0.4$~$\AA, EW([\ion{Fe}{xi}])=3.0$\pm$0.7$~$\AA$~$ and EW([\ion{Fe}{xiv}])=5.2$\pm$0.4$~$\AA. But the EW measurements of [\ion{Fe}{x}] and [\ion{Fe}{xi}] have not been galaxy-subtracted, they are therefore underestimated. 
Since no information  is given to estimate the contribution of the stellar component near these lines, we take EWs for [\ion{Fe}{x}] and [\ion{Fe}{xi}] of 4$~$\AA, estimating the effect of a dilution of about 1/3. This value is close to that found by Serote-Roos et al. ({\cite{Serote}) for \object{NGC 3516} which displays an EW of the Calcium triplet about equal to that of \object{MCG-6-30-15} (Morris $\&$ Ward \cite{Morris}). The [\ion{Fe}{xiv}] 5303\AA$~$ line is blended with [\ion{Ca}{v}] 5309\AA, so its EW is overestimated. That is why we also use another EW value for [\ion{Fe}{xiv}] of 3$~$\AA$~$ which should be closer to the real value.\\

\indent In Figs.~\ref{f11}, \ref{f12}, \ref{f13} and \ref{f14} both values obtained by Reynolds et al. (\cite{Reynolds95}) for each optical depth are displayed ($\tau_{\ion{O}{vii}}$=0.53 and 0.63: thick lower and upper solid lines respectively and  $\tau_{\ion{O}{viii}}$=0.19 and 0.44: thick lower and upper long dashed lines respectively). Isovalue curves for the optical iron coronal lines are also reported (thin lower and upper long dashed lines: EW([\ion{Fe}{x}])=3 and 4$~$\AA, thin lower and upper dot-dashed lines: EW([\ion{Fe}{xi}])=3 and 4$~$\AA, thin lower and upper solid lines: EW([\ion{Fe}{xiv}])=3 and 5$~$\AA$~$ respectively). 

\indent In order to reproduce both $\tau_{\ion{O}{vii}}$ and $\tau_{\ion{O}{viii}}$ values of July 1993 (one-zone model), an ionization parameter of the order of 200--400 is needed depending on the shape of the ionized spectrum and of the type of model (pure photoionized or hybrid). A range of $\xi$ between 300 to 900 is required for the August 1993 data. These values are significantly greater than values found by Reynolds et al. (\cite{Reynolds95}), reflecting the different shape of the ionization continuum (power law) they assumed. The ratio of the ionization parameter values which are derived for both epochs allows to get rid of the shape of the ionizing continuum. This ratio is consistent with the one obtained by Reynolds et al. ($\sim$1.3) except for the pure photoionized model with the ``AGN continuum'' ($\sim$2.6).\\
\indent However, the short variability time-scale ($\sim$10\,000 s) of the \ion{O}{viii} edge favors a two-zone model (Reynolds et al. \cite{Reynolds95}, Otani et al. \cite{Otani}).\\
\indent For the pure photoionization model with the ``AGN continuum'' a n$_{H}$ value of about 10$^{9}$ cm$^{-3}$ could account for the [\ion{Fe}{xiv}] (EW$\sim$3\AA) for a narrow range of $\xi$: 150--300 (cf Fig.~\ref{f12}).\\
\indent For the range of densities considered here, the inner zone responsible of the \ion{O}{viii} edge would contribute weakly to the coronal lines of \ion{Fe}{x} and \ion{Fe}{xi} and it is not constrained by the [\ion{Fe}{xiv}] line at very high values of $\xi$.\\
\indent The recombination time-scale derived for the high density value associated with the \ion{O}{vii} edge is much smaller than the variability time-scale of the associated region. Thus, the photoionization equilibrium can be applied.\\
\indent We also point out that our computations are restricted to dust-free models, whereas dust inside the outer warm absorber has been considered as a viable solution by Reynolds et al. (\cite{Reynolds97}).

\section{Conclusions} \label{sec:Conclusion}

\indent Using a photoionization code, including the most recent atomic data available for the coronal lines, we have found that the coronal lines could be formed in the Warm Absorber of the Seyfert 1 galaxies, and they strongly constrain its physical parameters, especially the hydrogen density. In order to take into account the available observational constraints, a high density is required for the mean observed Seyfert 1 features, as well as for the case of \object{MCG-6-30-15}, for both considered models (photoionized medium in or out of thermal equilibrium). 
A model with two different regions is favored, an inner zone mainly producing the \ion{O}{viii} associated with a high ionization parameter ($\xi\sim$a few hundreds) and an outer zone where the \ion{O}{vii} edge is formed, corresponding to a smaller $\xi$ ($\sim$a few tens).
A density n$_{H}\sim$10$^{10}$cm$^{-3}$ and $\xi\sim$10--100 for a typical Seyfert 1 \ion{O}{vii} edge implies a radius similar to that of the BLR (R$\sim$a few 10$^{16}$ cm). For higher $\xi$ producing the \ion{O}{viii} edge, a region of low density ($\sim$10$^{8}$ cm$^{-3}$) is not obviously ruled out, which would be at a similar distance as the BLR, while a more likely high density region ($\sim$10$^{10}$ cm$^{-3}$) would be located even inside the BLR.\\
\indent The gas pressure being proportional to the ratio of the temperature over the ionization parameter (P$_{gas}\propto~$T/$\xi$), we deduce that the pressure is the same in the BLR and in the WA (using T$_{BLR}\sim$10$^{4}$K and $\xi_{BLR}\sim$10, T$_{WA}\sim$10$^{5}$K and $\xi_{WA}\sim$100). So the WA could coexist with the BLR and be a second gaseous phase of this medium.\\
\indent In this paper, several assumptions such as a solar abundance, a constant density, a covering factor of 0.5 and a dust-free medium have been made. This analysis is restricted to available coronal and high-ionization resonance line measurements. In a near future additional constraints will be brought by measurements of coronal lines in the IR and of X-ray resonance lines which will be detectable thanks to the next generation of X-ray telescopes (AXAF, XMM, ASTRO-E). Our knowledge of the WA should also be improved by X-ray temporal variability studies. These investigations are crucial since they provide important diagnostics for the physical conditions which prevail in the ionized plasma. For instance, if the recombination time-scale is larger than the variability time-scale of the source, photoionization equilibrium could not be applied (Reynolds $\&$ Fabian \cite{Reynolds Fabian}).\\
\indent In the same way the fact that coronal lines could be formed in the WA should be confirmed by detection of rapid variations of these lines, which have not yet been observed.

\begin{acknowledgements}
We acknowledge Monique Joly for fruitful discussions and Claude Zeippen for helpful conversations about atomic processes.
\end{acknowledgements}


\begin{thebibliography}{}
\bibitem[1973]{Allen}
Allen, C. W., 1973, 
in ``Astrophysical quantities'', London: University of London, Athlone Press, 3rd ed., p31
\bibitem[1991]{Appenzeller}
Appenzeller, I., Wagner, S. J., 1991 A$\&$A 250, 57
\bibitem[1970]{Bely}
Bely, O. $\&$ Faucher, P., 1970, $A\&A$ 6, 88
\bibitem[1992]{Boller}
Boller, Th., Meurs, E. J. A., Brinkmann, W., et al., 1992, $A\&A$ 261, 57
\bibitem[1997]{Dere}
Dere, K. P., Landi, E., Mason, H. E. et al., 1997, A$\&$AS 125, 149
\bibitem[DP98]{Dumont}
Dumont, A.-M., Porquet, D., 1998, in preparation (DP98) 
\bibitem[E97]{Erkens}
Erkens, U., Appenzeller, I., Wagner, S., 1997, A$\&$A 323, 707 (E97)
\bibitem[1985]{Feigelson}
Feigelson, E. D., Nelson, P. I., 1985, ApJ 293, 192
\bibitem[1991]{Ferland91}
Ferland, G. J. 1991, Ohio State University, Astronomy Department Internal Report 91--01 
\bibitem[1998]{Ferland98}
Ferland, G. J., Korista, K. T., Verner, D. A. et al., 1998, PASP 110, 761
\bibitem[1995]{Genzel}
Genzel, R., Weitzel, L., Tacconi-Garman, L. E. et al., 1995, ApJ 444, 129 
\bibitem[1995]{George}
George, I. M., Turner, T. J., Netzer, H., 1995, ApJ 438, L67
\bibitem[1998]{George98}
George, I. M., Turner, T. J., Netzer, H. et al., 1998, ApJS 114, 73
\bibitem[1995]{Giannuzzo}
Giannuzzo, E., Rieke, G. H., Rieke, M. J., 1995, ApJ 446, L5
\bibitem[1978]{Grandi}
Grandi, S. A., 1978, ApJ 221, 501 
\bibitem[1996]{Guainazzi}
Guainazzi, M., Mihara, T, Otani, C., Matsuoka, M., 1996, PASJ 48, 781
\bibitem[1984]{Halpern}
Halpern, J. C., 1984, ApJ 281, 90
\bibitem[1995]{Hamann}
Hamann, F., Shields, J. C., Ferland, G. J., Korista, K. T., 1995, ApJ 454, 688
\bibitem[1983]{Heil}
Heil, T. G., Kirby, K., Dalgarno, A., 1983, Phys. Rev. A 27, 2826 
\bibitem[1989]{Korista}
Korista, K. T., Ferland, G. J., 1989, ApJ 343, 678 
\bibitem[1995]{Krolik95}
Krolik, J. H.,  Kriss, G. A., 1995, ApJ 447, 512
\bibitem[1980]{Landman}
Landman, D. A., 1980, ApJ 240, 709
\bibitem[1997]{Laor}
Laor, A., Fiore, F., Elvis, M. et al., 1997, ApJ 477, 93
\bibitem[1992]{Lavalley}
Lavalley, M., Isobe, T., Feigelson, E., 1992 in  ``Astronomical Data Analysis Software and Systems I'', A.S.P. Conference Series Vol. 25, Diana M. Worrall, Chris Biemesderfer, and Jeannette Barnes, eds., p. 245.
\bibitem[1975]{Mason}
Mason, H. E., 1975, MNRAS 170, 651
\bibitem[1987]{Mathews}
Mathews, W. G. $\&$ Ferland, G.J., 1987 ApJ 323, 456 
\bibitem[1994]{Mihara}
Mihara, T., Matsuoka, M., Mushotzky, R. F. et al., 1994, PASJ 446, L137
\bibitem[1991]{Moorwood}
Moorwood, A. F. M., Oliva, E., 1991, The Messenger 63, 57
\bibitem[1988]{Morris}
Morris, S. L., Ward, M. J., 1988, MNRAS 230, 639
\bibitem[1994]{Nandra}
Nandra, K., Pounds, K. A., 1994, MNRAS 268, 405
\bibitem[1993]{Netzer93}
Netzer, H., 1993, ApJ 411, 594 
\bibitem[1996]{Netzer96}
Netzer, H., 1996, ApJ 473, 781
\bibitem[1970]{Nussbaumer}
Nussbaumer, H., $\&$ Osterbrock, D.E., 1970, ApJ 161, 811 
\bibitem[1968]{Oke}
Oke, S., Sargent, W., 1968, ApJ 151, 807 
\bibitem[1994]{Oliva}
Oliva, E., Salvati, M., Moorwood, A. F. M., Marconi, A., 1994, A$\&$A 288, 457 
\bibitem[1997]{Orr}
Orr, A., Molendi, S., Fiore, F. et al., 1997, A$\&$A 324, L77
\bibitem[1969]{Osterbrock}
Osterbrock, D. E., 1969, Astrophys. Letters 4, 57
\bibitem[1996]{Otani}
Otani, C., Kii, T., Reynolds, C. S. et al., 1996, PASJ 48, 211  
\bibitem[1995]{Pelan}
Pelan  J., Berrington, K. A., 1995, $A\&A$ 110, 209
\bibitem[1984]{Penston}
Penston, M. V., Fosbury, R. A. E., Boksenberg, A. et al., 1984, MNRAS 208, 347
\bibitem[1997]{Piro}
Piro, L., Balucinska-Church, M., Fink, H. et al., 1997, $A\&A$ 319, 74
\bibitem[1998]{Porquet}
Porquet, D., Dumont, A.-M., 1998, in "Structure and Kinematics of Quasar Broad Line Regions", Eds C. M. Gaskell, W. N. Brandt, M. Dietrich, D. Dultzin-Hacyan, and M. Eracleous, ASP Conf. Ser., in press
\bibitem[R97]{Reynolds}
Reynolds, C. S., 1997, MNRAS 286, 513 
\bibitem[1995]{Reynolds Fabian}
Reynolds, C. S., Fabian, A. C., 1995, MNRAS 273, 1167
\bibitem[1995]{Reynolds95}
Reynolds, C. S., Fabian, A. C., Nandra, K. et al., 1995, MNRAS 277, 901
\bibitem[1997]{Reynolds97}
Reynolds, C. S., Ward, M. J., Fabian, A. C., Celotti, A., 1997, MNRAS 291, 403
\bibitem[1996]{Rush}
Rush, B., Malkan, M. A., Fink, H. H., Voges, W., 1996, ApJ 471, 190
\bibitem[1997]{Schartel}
Schartel, N., Schmidt, M., Fink, H. H. et al., 1997, $A\&$A 320, 696 
\bibitem[1996]{Serote}
Serote-Roos,M., Boisson, C., Joly, M., Ward, M. J., 1996, MNRAS 278, 897
\bibitem[1995]{Shields}
Shields, J. C., Ferland, G. J., Peterson, B.M., 1995, ApJ 441, 507
\bibitem[1992]{Spinoglio}
Spinoglio, L., Malkan, M.A., 1992, ApJ 399,504  
\bibitem[1996]{Storey}
Storey, P. J., Mason, H. E., Saraph, H. E., 1996, $A\&$A 309, 677
\bibitem[1996]{Thompson96}
Thompson, R.I., 1996, ApJ 459, L61 
\bibitem[1989]{Viegas-Aldrovandi}
Viegas-Aldrovandi, S. M., Contini, M., 1989, A$\&$A 215, 253 
\bibitem[1995]{Zheng95}
Zheng, W., Kriss, G. A., Davidsen,A. F., 1995, ApJ 440, 606 
\bibitem[1997]{Zheng97}
Zheng, W., Kriss, G. A., Telfer, R. C. et al., 1997, ApJ 475, 469 
\end{thebibliography}
\end{document}